# Connecting the dots across time: Reconstruction of single cell signaling trajectories using time-stamped data

Sayak Mukherjee[1,8], David Stewart[6], William Stewart[1,4], Lewis L. Lanier[7], Jayajit Das*[1,2,3,5]

[1]Battelle Center for Mathematical Medicine, Research Institute at the Nationwide Children's Hospital, 700 Children's Drive, OH 43205
Departments of [2]Pediatrics, [3]Physics, [4]Statistics, and the [5]Biophysics Program, the Ohio State University, Columbus, OH
[6]Department of Mathematics, University of Iowa, Iowa City, IA
[7]Department of Microbiology and Immunology, University of California, San Francisco, San Francisco, CA 94143
[8] Present address: Institute of Bioinformatics and Applied Biotechnology, Electronic City Phase I, Bangalore, India 560100

Single cell responses are shaped by the geometry of signaling kinetic trajectories carved in a multidimensional space spanned by signaling protein abundances. It is however challenging to assay large number (>3) of signaling species in live-cell imaging which makes it difficult to probe single cell signaling kinetic trajectories in large dimensions. Flow and mass cytometry techniques can measure a large number (4 - >40) of signaling species but are unable to track single cells. Thus cytometry experiments provide detailed time stamped snapshots of single cell signaling kinetics. Is it possible to use the time-stamped cytometry data to reconstruct single cell signaling trajectories? Borrowing concepts of conserved and slow variables from non-equilibrium statistical physics we develop an approach to reconstruct signaling trajectories using snapshot data by creating new variables that remain invariant or vary slowly during the signaling kinetics. We apply this approach to reconstruct trajectories using snapshot data obtained from *in silico* simulations, live-cell imaging measurements, and, synthetic flow cytometry datasets. The use of invariants and slow variables to reconstruct trajectories provides a radically different way to track object using snapshot data. The approach is likely to have implications for solving matching problems in a wide range of disciplines.

# 1. Introduction

Tracking signaling events in single cells is a key step towards understanding single cell response mechanisms. Signaling events are orchestrated by a large number of intercellular molecular species that transmit information pertaining to changes in the extracelluar environment to the cell nucleus(1). Single cell responses are often influenced by the geometry of multidimensional temporal trajectories describing time evolution of single cell protein abundances. For example, in human cancer cell lines, fold change in the abundance of the protein phosphorylated Erk (or pErk) as opposed to the absolute value of pErk abundance regulates single cell growth responses(2). In general, the signaling kinetics of many molecular species could affect cell decision processes. However, our understanding regarding the link between the geometry of signaling kinetic trajectories and single cell responses are often limited to few (1-3) molecular species. This is because spectral overlap between fluorescent dyes, and, photo-toxicity induced by excited fluorophores(3) make it challenging to track a large number (>3) of molecular species in live cell imaging experiments. Flow cytometry (4, 5) and recently developed mass cytometry experiments (4, 5) can assay 4 - >40 proteins simultaneously in single cells at multiple times, but it is not possible to track single cells in these experiments. Is it possible to reconstruct single cell trajectories, even approximately, from such time-stamped snapshot data? An affirmative answer to this question will be valuable for analyzing signaling mechanisms or calculation of autocorrelation functions(6) for extracting relevant time scales and inference of signaling networks.

Tracking multiple objects across time using time-stamped data is a common problem encountered in diverse research areas ranging from tracking single molecules in live cells(7) to fluid particles in turbulent flows(8) to birds flying in a flock(9) to tracking individuals in surveillance videos(10). The difficulty in tracking individual objects in these problems is characterized by a dimensionless parameter $\xi = \Delta l/l_0$, where, $\Delta l$ is the average distance an object moves between two successive time recordings and $l_0$ $(=\rho^{-1/d})$ is the average object-to-object distance for the objects distributed in $d$ dimensions with a density $\rho$(8, 11) (Fig. 1). When, $\xi \ll 1$, connecting objects across time is straightforward, whereas, when $\xi \gg 1$, matching objects across time becomes ambiguous. In the later scenario, two types of methods are used to generate solutions - either by optimizing cost functions that are often constructed in an empirical manner (8), or by evaluating probabilities for different matching configurations by estimating parameters in an underlying model(11). The success of the first method depends on stumbling upon a good cost function and the second method requires intensive computation for the evaluation of the likelihood function and estimation of the model parameters.

Signaling processes usually involve a large number of molecular species (tens to hundreds). Depending on the available antibodies cytometry techniques assay 4 - >40 molecular species in $10^3$-$10^6$ cells at chosen time points where the abundances of the marker species have changed appreciably (12). Thus, the time-stamped data collected in cytometry experiment are almost always in the $\xi \gg 1$ range. In addition, involvement of a large number of molecular species in generating a signaling response demands analysis of the kinetic trajectories in high dimensions (d>3). This can create computational road blocks regarding parameter estimation of an underlying model or particle filters using Markov Chain Monte Carlo (MCMC)-based algorithms(11, 13), which have been extensively used for tracking fluid particles (spatial dimension, d=2) or flocking birds (spatial dimension, d=3). The reconstruction using cytometry data is further complicated as, unlike the above examples, the same objects (i.e., single cells) are not assayed at successive time points.

A fundamentally different way of approaching this problem would be to use the measured variables to construct new variables ($I$ in Fig. 1) that do not change ($\Delta l^{(I)} = 0$) or change more slowly ($\Delta l^{(I)} \to 0$) with time, while still varying appreciably between the objects at a fixed time point. Thus, when the density of the objects ($\rho^{(I)}$) in the $d_{(I)}$ dimensional space of the new variables $I$ is small (i.e., large $l_0^{(I)} = (1/\rho^{(I)})^{1/d_I}$), the matching problem posed in terms of $I$ will result in a substantial reduction in the parameter $\xi^{(I)}$ ($= \Delta l^{(I)}/l_0^{(I)}$), which can even fall in the range, $\xi^{(I)} \ll 1$, where tracking objects is straightforward. In addition, when $\xi \gg \xi^{(I)} > 1$, tracking objects in $I$ becomes more amenable to the standard techniques(11, 13) due to the lower dimensionality of the manifold and the smaller value of $\xi^{(I)}$. However, it is often difficult to construct such conserved or slow variables for the problem in hand. In physical systems, conservation laws (e.g., conservation of energy)(14, 15), break down of continuous symmetry (e.g., Goldstone modes)(14, 15), or, the presence of a critical point (14, 15) can bring conserved and slow variables into existence. However, direct application of such principles in cell signaling processes is not obvious.

Here we developed a framework for matching untagged single cells across time by constructing conserved and slow variables using abundances of molecular species that follow biochemical signaling kinetics. The calculations of the new variables do not require any parameter estimation, and thus avoid computational difficulties usually encountered in matching problems. We constructed two invariants for an ideal kinetics where the signaling kinetics is described by a closed system of first order reactions. One of the invariant variables is based on the conservation of total number of molecular species, and, the other one involves scaling the measured abundances by a particular function of the covariance matrix. These invariants turn into slow variables or remain as invariants for more general biochemical reactions that include external fluxes, involve second or higher order reactions, and contain stochastic fluctuations (16). The slow variables and invariants, constructed from the measured species abundances, allowed us to connect a single cell with a "sister cell" at a later time whose species abundances are reasonably close to the correct cell partner. We validate the above formalism by reconstructing trajectories using published live cell imaging data (17). Lastly, we apply our framework for reconstructing signaling trajectories between successive time points in a synthetic flow cytometry dataset.

## 2. Results

### 2.1. Development of the framework
In this section we describe the development of the framework and then evaluate the framework on a range of signaling models based on deterministic linear, deterministic non-linear, and, stochastic reactions.

### A. Determination of invariants in an ideal system

We constructed two invariants, $I_T$ and $I_M$, for an ideal system of biochemical reactions. The invariants do not change with time but vary between single cells at a particular time. The ideal system satisfies the following conditions: (i) the reaction propensities are linear functions of the copy numbers (or abundances) of the molecular species, (ii) the reaction rates are time independent and there is no external in-(or out-)flux of molecules, and, (iii) the kinetics does not include any intrinsic noise fluctuations. The mass action kinetics of the biochemical reactions are described as an autonomous system of linear first order ordinary differential equations (ODEs). Consider an ensemble of $N$ number of single cells where a single cell (indexed by $\alpha$) contains $n$ different molecular species (indexed by $i$) with abundances $\{x_i^{(\alpha)}\}$. Any pair of molecular species, $i$ and $j$, in an individual cell (say, the $\alpha^{th}$ cell) react following the above conditions, and, the propensity for the reaction $j \rightarrow i$ (or, $i \rightarrow j$) is given by $M_i^j x_j^{(\alpha)}$ (or $M_j^i x_i^{(\alpha)}$). The reaction rates $M_i^j$ and $M_j^i$ are always positive and constant as long as, $i \neq j$. In our notation scheme the $(i,j)$ element of the M matrix, $M_i^j$, is associated with a reaction, $j$ (superscript index)$\rightarrow i$ (subscript index). Vanishing values for both $M_i^j$ and $M_j^i$ would imply absence of any reaction between the species $i$ and $j$. In addition, the elements of the M matrix do not depend on the cell index implying that the signaling reactions occur with the same rates in individual cells.

The species abundances in individual cells follow a mass-action linear kinetics described by a set of coupled first order linear ODEs,

$$\frac{dx_i^{(\alpha)}}{dt} = \sum_{j=1}^{n} M_i^j x_j^{(\alpha)} \qquad (1)$$

Since the elements of the M matrix do not depend on time explicitly, Eq. (1) represents an autonomous system(18).

The source of variations in species abundances following the kinetics in Eq. (1) are the cell-to-cell variations in the pre-stimulus condition ($\{x_i^{(\alpha)}(t=0)\}$) arising due to cell-cell differences in total content of the molecular species and tonic (basal) signaling(19). These variations are known as extrinsic noise fluctuations (16, 20). In the ideal case we assume that cytometry experiments can measure all the signaling species abundances in Eq. (1) at any time. Below we describe the invariants, $I_T$ and $I_M$.

(i) *Total abundance* ($I_T$). If the rates in Eq. (1) are further constrained to satisfy, $\sum_{i=1}^{n} M_i^j = 0$, then the total copy number of the molecular species 1 to n remains unchanged over time in a single cell #α , i.e.,

$$I_T^{(\alpha)} = \sum_{i=1}^{n} x_i^{(\alpha)}(t) = \sum_{i=1}^{n} x_i^{(\alpha)}(0) \qquad (2)$$

An example of the above case could be first order reactions describing phosphorylation and dephosphorylation of a single protein species preserving the total number of protein molecules. The number conservation is an elementary conservation principle followed by biochemical reactions of first or higher order reactions in the absence of any particle exchange with the environment. We will analyze implications of this conservation principle for non-ideal cases in later sections.

(ii) *Magnitude of the scaled abundance vector* ($I_M$). The magnitude of the scaled abundance vector $\overrightarrow{\tilde{x}^{(\alpha)}} \equiv \{\tilde{x}_1^{(\alpha)}, \tilde{x}_2^{(\alpha)}, \cdots, \tilde{x}_n^{(\alpha)}\}$ in a single cell (#α) remains unchanged throughout the kinetics from its value at t=0. $\tilde{x}_i^{(\alpha)}$ is defined as,

$$\tilde{x}_i^{(\alpha)}(t_1) = \sum_{j=1}^{n} \left[ (J(t_1))^{-1/2} \right]_{ij} x_j^{(\alpha)}(t_1) \qquad (3)$$

The derivation is shown in the Materials and Methods section. The elements of the J matrix ($\{J_{ij}\}$) in Eq. (3) denote the covariances of the species abundances at any time t

$$J_{ij}(t) = \frac{1}{N}\sum_{\alpha=1}^{N}(x_i^{(\alpha)}(t)-\mu_i(t))(x_j^{(\alpha)}(t)-\mu_j(t)) \quad (4)$$

and, $\{\mu_i\}$ are the mean species abundances,

$$\mu_i(t) = 1/N\sum_{\alpha=1}^{N} x_i^{(\alpha)}(t). \quad (5)$$

Both $\{\mu_i\}$ and $\{J_{ij}\}$ can be calculated from the cytometry snapshot data.

The magnitude of $\vec{\tilde{x}}^{(\alpha)}$ defined as,

$$m^{(\alpha)}(t) = \left|\vec{\tilde{x}}^{(\alpha)}(t)\right| = \sqrt{\sum_{i=1}^{n}\left(\tilde{x}_i^{(\alpha)}(t)\right)^2}$$

does not change with time, i.e.,

$$I_M^{(\alpha)} = m^{(\alpha)}(t_1) = m^{(\alpha)}(t_2) = m^{(\alpha)}(t=0) \quad (6)$$

is an invariant of the kinetics. One can physically interpret the transformation in Eq. (3) following way. The first order chemical reactions rotate (or reflect) as well as stretch (or compress) the abundance vector $\vec{x}^{(\alpha)}(t)$ with time. The transformation in Eq. (3) rescales the vector to offset the stretching (or compressing) component. Subsequently, the time evolution of the scaled vector can be described as a pure rotation (or reflection) (see web supplement). Since rotation or reflection is an orthogonal transformation, the magnitude of the scaled variable is preserved.

*Exact matching using invariants:* $I_T$ and $I_M$ are functions of copy numbers of molecular species, and, therefore vary considerably from cell to cell at a given time point. Thus, these invariants create unique "tags" for single cells, and, pairing single cells across time then is reduced to matching the same values of invariants in cell populations at two time points. A possible degeneracy can arise when an invariant takes the same value in multiple single cells. For example, single cells (e.g., #α and #β) with abundances lying in the plane defined $I^{(\alpha)}_T = \Sigma_i x^{(\alpha)}_i(0) = \Sigma_i x^{(\beta)}_i(0) = I^{(\beta)}_T$ cannot be resolved by $I_T$. In such cases the invariant $I^{(\alpha)}_T$ will be unable to match these cells across time. This same difficulty holds for the other invariant $I^{(\alpha)}_M$, however, it is less likely to encounter such degeneracy in this case.

## B. Construction of slow variables for non ideal situations

Cell signaling networks often deviate from the ideal kinetics considered above. This occurs due to multiple reasons: (i) It is usually infeasible to assay all the molecular species involved in a signaling network. In that case, the measured species abundances evolve as a non-autonomous system because the unmeasured species abundances implicitly give rise to time-dependent reaction rates or external fluxes in the kinetics. (ii) The propensities of biochemical signaling reactions are often nonlinear functions of the species abundances. (iii) The copy numbers of molecular species contain stochastic fluctuations arising from intrinsic noise in biochemical reactions. In the presence of the above effects, except few special cases, both $I_T$ and $I_M$ cease to behave as invariants of the kinetics. Our investigations (details in subsection C) using simulations of a variety of *in silico* signaling kinetics showed that for specific subsets of species abundances, the variables $I_M$ and $I_T$, evolve at a much slower rate compared to the measured species abundances in a time interval. Borrowing from the lexicon of non-equilibrium statistical physics(15), we designate these variables as "slow variables". Next, we discuss our scheme to identify appropriate combination of measured species abundances where $I_T$ and $I_M$ behave as slow variables or invariants (also see Fig. 2).

We used a metric known as the Jensen-Shannon divergence JSD ($0 \leq$ JSD $\leq \ln2$)(21, 22) to determine slow variables and invariants in our scheme. JSD characterizes the difference between a pair of distributions(22). JSD vanishes when distributions are identical and increases monotonically as the overlap between the distributions decreases. JSD is defined as,

$$JSD^{(y)} = \frac{1}{2}\left[d_{KL}(P_1(y) \| M) + d_{KL}(P_2(y) \| M)\right] \quad (7)$$

where, M(y)=1/2(P$_1$(y)+ P$_2$(y)). The *y* superscript in JSD$^{(y)}$ denotes the variable in the distributions. d$_{KL}$ is the Kullback-Leibler divergence (21) between two distributions. When y is a discrete variable, d$_{KL}$ is given by,

$$d_{KL}(P_1(y) \| M) = \sum_y P_1(y) \ln\left[P_1(y) / M(y)\right]$$

When I ∈ ($I_T$,$I_M$) is an invariant of the kinetics, the distribution of *I* in a cell population does not change with time, i.e.,

P(I,$t_1$)=P(I,$t_2$) . (8)

An exception to this can arise where *I* varies between individual cells yet the distribution of *I* remains unchanged across time. This can occur when the stochastic kinetics of the chemical reactions is in the steady state where species abundances (and *I*) change across time in individual cells but the distributions of these variables remain time independent. In this situation, the presence of the equality as in Eq. (8) will not imply an existence of a slow variable, *I*. However, such situations can be easily detected from the data by checking if distributions of original variables do not vary across time. Therefore, the JSD between P(*I*,$t_1$) and P(*I*,$t_2$) vanishes when *I* is an invariant, *i.e.*, $JSD^{(I)}$=0. When *I* does not behave as an invariant, then, $JSD^{(I)}$>0 (Fig. 2). However, I ∈ {$I_T$, $I_M$} can still evolve at a slower rate than any of the measured species in the time interval $t_1$ to $t_2$. We evaluated this possibility by comparing changes in the JSD values corresponding to the distributions of species abundances and *I* in a cell population in the time interval $t_1$ to $t_2$. When $I_T$ or $I_M$, evolves slower than any of the individual species (say x) abundances in set of species, specifically, when

$$JSD^{(I_T \text{ or } I_M)} < JSD^{(x)}$$
(9)

, we denote I ($I_T$ or $I_M$) as the slow variable for that set of species abundances. In general, one or more of the species abundances in a set of all the measured species will not satisfy the condition in Eq. (9), therefore, $I_T$ or $I_M$ cannot be considered as a slow variable for that set. However, for a subset of the measured abundances, $I_T$ or $I_M$, could still behave as a slow variable or an invariant. To explore this possibility, we compared the JSD values corresponding to species abundances and *I* ∈ ($I_M$, $I_T$) in a time interval $t_1$ - $t_2$ for all possible subsets that can be constructed using the measured abundances (Fig. 2). We used a classification scheme for the subsets (Fig. 2B) to describe our results. When *n* number of species abundances are measured, we considered all possible combinations ($2^n$-n-1) of the abundances excluding the singletons.

These subsets were organized into classes where each class (indexed by *k*) contained subsets with the same cardinality (*k*) (Fig. 2B). The cardinality is defined as the number of species abundances in a subset, e.g. for *n*=14 the class with cardinality *k*=4 (or class

#4) contains $^{14}C_4 = 1001$ different subsets with each subset composed of 4 different molecular species abundances. In addition, the subsets within class #$k$ were indexed by the integers ({$m$}) 1 to $^nC_k$. Thus, a particular subset is denoted by a class number $k$ and a subset index $m$ (Fig. 2B).

*Matching using slow variables and invariants:* We created a cost function $E$, that measures the total Euclidean distance between slow variables in pairs of single cells ({α} at $t_1$, {β} at $t_2$) across time. E is defined as,

$$E(\{\beta\}) = \sum_{\alpha=1}^{N} \left( I^{(\alpha)}(t_1) - I^{(\beta)}(t_2) \right)^2 \qquad (10)$$

The "sister cells" constitute the set {$β^M$} that minimizes E. The minimization amounts to a bipartite matching between the set {α} and {β}, we used an algorithm (see Methods) based on sorting with an O(n$ln$(n)) computation time. A sister cell can be thought of as a partnering cell whose species abundances are not substantially different than that in the correct cell partner.

*Quantification of quality of matching*: We calculated the error in the reconstructed trajectory when a cell α at time $t_1$ was paired with a cell β (or the "sister cell") at a later time point $t_2$ instead of the correct partner cell β' at $t_2$. β' is uniquely determined by α for deterministic signaling kinetics in the non-steady state. We defined an error $\chi_{\alpha\beta}$ for pairing cell #α at time $t_1$ with cell #β at time $t_2$:

$$\chi_{\alpha\beta} = \sqrt{\sum_{i=1}^{N} \left( x_i^{(\beta)} - x_i^{(\beta')} \right)^2} \qquad (11)$$

$\chi_{\alpha\beta} = 0$, when β = β'. Note, this metric identifies the cells by the measured species abundances in individual cells, e.g., if the sister cell possesses the same values of the measured abundances as the correct cell partner, the sister cell is identified as the correct match. It is possible that the sister cell contains different values for unmeasured species abundances compared to the correct cell; however, Eq. (11) remains blind to such differences. A small $\chi_{\alpha\beta}$ would imply a small difference between the correct partner (β') and the "sister cell" (β) for the measured species in the subset. We calculated a distribution of $\chi_{\alpha\beta}$ (P(χ)) for the cell pairs matched using our scheme and compared that with the case when cells were paired randomly across time.

We also calculated the auto-correlation function, $A_{ij}(t_1, t_2)$, between species $i$ at time $t_1$ and species $j$ at time $t_2$ for the pairings with the correct partner cells and the sister cells. $A_{ij}(t_1, t_2)$ between the correct cell pairs ($\{\#\alpha$ paired with $\#\beta'\}$) is defined as,

$$A_{ij}^{correct}(t_1,t_2) = \frac{1}{N}\sum_{\alpha=1}^{N}(x_i^{(\alpha)}(t_1)-\mu_i(t_1))(x_j^{(\beta')}(t_2)-\mu_j(t_2)) \quad (12a)$$

This is compared with the autocorrelation function corresponding to the pairings with the sister cells,

$$A_{ij}^{sister}(t_1,t_2) = \frac{1}{N}\sum_{\alpha=1}^{N}(x_i^{(\alpha)}(t_1)-\mu_i(t_1))(x_j^{(\beta(\alpha))}(t_2)-\mu_j(t_2)) \quad (12b)$$

where the a sister cell for $\alpha$ is indexed by $\beta(\alpha)$. The difference between the autocorrelations between the pairings with the correct (Eq. (12a)) and the sister cells (Eq. (12b)) is calculated by distance $\Delta(A^{correct}, A^{sister})$ between the matrices:

$$\Delta(A^{correct},A^{sister}) = \sqrt{\sum_{i=1}^{n}\sum_{j=1}^{n}\left|A_{ij}^{correct}-A_{ij}^{sister}\right|^2} \quad (13)$$

## C. Evaluation of slow modes and quality of matching for model signaling kinetics

Here we investigated the occurrence of slow modes ($I_M$ and $I_T$) in subsets of molecular species involved in model biochemical signaling networks as a proof-of-concept for the framework we developed in subsections A and B. We studied deterministic and stochastic kinetics in a system of first order reactions, and, the Ras activation signaling network composed of non-linear biochemical reactions (23). We assumed that all the signaling species as measured species for the *in silico* networks.

**(i) Deterministic first order kinetics.** We studied the matching problem for a signaling kinetics described by first order reaction kinetics composed of 14 different species (Fig. S1). The ODEs describing the mass action kinetics of all the 14 species is autonomous, however, when subsets of the 14 species are considered, the kinetics is no longer autonomous because the corresponding ODEs contain time-dependent external fluxes arising from the implicit kinetics of the unmeasured species.

*Identification of slow variables and invariants*: We analyzed $2^{14}-14-1= 16369$ different subsets in a time interval where the kinetics is not in the steady state. We acknowledge that the number of possible subsets can become prohibitively large at very large dimensions. We found that for every class ($k \geq 2$), the subsets produced a wide range of ($0 \leq JSD^{(I)} < \ln(2)$) $JSD^{(I)}$ values (Fig. 3A) corresponding to $I_T$ (Fig. S2) and $I_M$ (Fig. 3A). Next we analyzed if $I_T$ or $I_M$ behaved as slow variables or invariants in the subsets that are associated with smaller values of $JSD^{(I)}$. The comparison of the minimum values of $JSD^{(I_M)}$ with JSD values for the fastest and slowest species abundances (Fig. 3B) in the subsets associated with minimum $JSD^{(I_M)}$ showed that $I_M$ evolved as a slow variable in most of those subsets. $I_M$ turned into an invariant when all the 14 species were included in the set (k=14). Similar behavior was found for $I_T$ as well (Fig. S3). The composition of the subsets associated with the minimum values of $JSD^{(I_M \text{ or } I_T)}$ depends on the variable type ($I_T$ or $I_M$) and the time interval, as well as on the rate constants of the reaction network.

*Matching using the slow variables:* The cost function in Eq. (10) was minimized using the subsets associated with minimum $JSD^{(I)}$ to find the sister cells. The quality of matching was evaluated by calculating the error in matching $\chi$ (Eq. (11)) for each single cell-sister cell pair. We show the results for the pairing carried out using $I_M$ here. $I_M$ is an invariant for the subset corresponding to k=14, and minimizing E produced an exact matching in that case(Fig. 3C). $I_M$ turned out to be slow variable for the other subsets and the quality of matching using $I_M$ was reasonable (Fig. 3C). The mean $\chi$ values ($\overline{\chi}$) for these pairings were 0.6- to 0.8-times lower than that for random pairing (Fig. 3C). Next, we calculated the autocorrelation function for the matched pairs. Similar to $\overline{\chi}$, errors in the autocorrelation function ranged from very small to moderate values (0.2- to 0.8-times less) compared to that when the single cells were paired randomly (Fig. 3D). The subsets with smaller number of species show lower error values (Fig. 3D), this could arise due to the sensitivity of the autocorrelation function to small errors in matching in higher dimensions. The quality of matching using $I_T$ was qualitatively similar to that of $I_M$ (Fig. S4). Interestingly, pairing using $I_T$ produced better agreement with the correct trajectories even when $I_T$ evolved faster than $I_M$ (Fig. S5). This behavior emphasized the role of the cost function in pair matching (also see the Discussion section).

**(ii) Nonlinear deterministic kinetics.** We used an experimentally validated signaling network for Ras activation in T lymphocytes (Fig. S6) (23). The network describes enzymatic activation of Ras by two enzymes SOS and Rasgrp1, where SOS-mediated Ras activation contains a positive feedback, i.e., an activated form of Ras or RasGTP induces a higher catalytic rate involving SOS. The deterministic kinetics displays bistability for a range of SOS and Rasgrp1 concentrations. We analyzed the non-steady state kinetics in the model for the parameter values that generate monostable or bistable steady states. The *in silico* model contained 14 different molecular species.

*Identification of slow variables and invariants*: Similar to the first order kinetics in the previous section, the $JSD^{(I)}$ values corresponding to $I_T$ showed large variations across subsets(Fig. S7). However, a noted difference from the previous example was $I_T$ behaved as an invariant for multiple subsets. This is because, even with the nonlinear rates, the total number of certain species is an invariant of the kinetics. For example, the total amount of Rasgrp1 contained in the complexes (free Rasgrp1, Rasgrp1-DAG, Rasgrp1-DAG-RasGDP) remained fixed throughout the kinetics, and, the subset {Rasgrp1, Rasgrp1-DAG, Rasgrp1-DAG-RasGDP} produced the minimum $JSD^{(I_T)}$ (=0) in the class k=3(Fig. 4A). $I_M$ in contrast, evolved only as a slow variable in specific subsets (Fig. S8).

*Matching using the slow variables:* For the subsets where $I_T$ behaved as an invariant, the matching produced exact alignment. For other subsets corresponding the minimum values of $JSD^{(I_T)}$, the pairing was substantially better than that of random matching (Fig. 4B). $I_M$-guided pairing showed small errors in abundances in the sister cells (Fig. S9); however, overall pairing using $I_T$ produced smaller errors.

**(iii) Stochastic signaling kinetics.** We simulated stochastic biochemical reactions in the first order reaction kinetics and Ras activation signaling kinetics by including intrinsic noise fluctuations. The simulations contained the same variations in the initial abundances used in the investigations for the deterministic kinetics. $I_T$ evolved as an invariant in subsets that were associated with conservation of the total number of molecular species (Fig. S10). $I_T$ also behaved as a slow variable in specific subsets (Fig. S10A). In contrast, $I_M$ behaved as a slow variable for select subsets (Fig. S10B). Pairing the single cells using $I_T$ for the subsets associated with minimum values $JSD^{(I_T)}$ showed reasonably smaller errors in the auto-correlation function compared to that for random pairings for most of the subsets (Fig. 4C). The matching using $I_M$ showed smaller errors (or $\chi$) compared to random pairing in select subsets (Fig. S11).

**2.2. Application of the framework for matching in live cell imaging data**

We reconstructed single cell gene expression kinetic trajectories by applying our framework using live-cell imaging data(17). Data were collected in single yeast cells where the transcription factor Msn2 translocated to the nucleus upon inhibition of protein kinase A (PKA) by a small molecule, 1-NM-PP1, and activated target fluorescent reporter genes CFP and YFP residing on homologous chromosomes in the diploid yeast cells (Fig. 5A). For our reconstruction we chose the kinetics of the dual reporter of the gene DCS2 (one of the seven genes activated in the study) induced by a single 40 minutes pulse of 1-NM-PP1 at a concentration of 690nM. The activation kinetics of the reporters depended non-linearly on the Msn2 abundance and also contained intrinsic and extrinsic noise fluctuations(17). We carried out our reconstruction method by treating the live-cell imaging data as snapshot data because it allowed us to compare the reconstructed trajectories with the measured single cell trajectories. We analyzed the simultaneous CFP and YFP gene expression kinetics data in 159 single cell trajectories(17). At each time point, we removed the single cell tag from the CFP and YFP expressions data and treated the 159 data points as snapshot data to perform the reconstruction. We reconstructed two-dimensional (CFP and YFP) single cell trajectories using either $I_T$ (Fig. 5B) or $I_M$ (Fig. S12), and, both showed a similar level of agreement between the measured and the reconstructed trajectories. We further quantified the quality of alignment using P($\chi$) (Fig. 5C) and autocorrelation functions (Fig. 5D, S12) for each reconstruction. Both indicators revealed lower errors in the reconstruction using *I* compared to random pairing.

## 2.3. Application of the framework for matching single cells using synthetic flow cytometry data

We applied our framework on synthetic flow cytometry data for matching single cells assayed at two successive time points. We used *in silico* or synthetic data instead of measured data from flow cytometry experiments for two main reasons: (i) The synthetic data allowed us to assess the quality of the reconstruction because the correct trajectories are known in the simulations. (ii) We were able to use species abundances in the synthetic data to calculate the slow variables. Single cell protein abundances are not directly accessible from the fluorescent intensities measured in flow cytometry. However, for many proteins it is possible to quantify single cell species abundances by calibrating the intensities against a standard curve (24). The synthetic flow cytometry data were generated at multiple time points by stochastic simulation (details in Materials and Methods) of the Ras activation network. The species abundances at the unstimulated state (or t=0) in individual cells were chosen from a multivariate normal distribution (Materials and Methods) to represent cell-cell differences in total protein abundances and tonic signaling. Six molecular species (RasGTP, RasGDP, SOS, RasGRP1, DAG, and, RasGAP) out of the 14 signaling species involved in the Ras signaling network (Fig. S6) were recorded in the synthetic data. We used different sets of cell populations to record single cell species abundances at any two time points in the synthetic data as the same cell populations are not assayed more than once in flow and mass cytometry experiments. We used $I_T$ instead of $I_M$ for all our reconstructions here because our previous investigations showed $I_T$ generated less error in reconstruction compared to $I_M$ for the Ras activation network. In addition to carrying out reconstruction of trajectories between a pair of successive time points we addressed the following relevant issues. (i) Is there a specific subset of measured abundances that produces a more accurate reconstruction compared to the other subsets? (ii) How does the error in the reconstruction depend on the size of the time interval between two successive measurements? (iii) How does the reconstruction using $I_T$ compare against other available methods for matching?

In our investigation we divided the six measured species into five classes (see Fig. 2 and section 2.1B for more details) composed of 2, 3, 4, 5, and, 6 species. In each class, we determined the subsets that produced the largest and the smallest $JSD^{(I_T)}$ values. The subset corresponding to the smallest $JSD^{(I_T)}$ gives rise to the slowest $I_T$ in a class. The subsets corresponding to the lowest $JSD^{(I_T)}$ values in a time interval also produced the lowest errors in the reconstruction in each class (Fig. 6A). All the reconstructions corresponding to the lowest $JSD^{(I_T)}$ values faired better than the random pairings (Fig. 6B). These results demonstrate that it is possible to identify a group of species in the cytometry dataset that will generate better reconstruction in a time interval. Moreover, when the interest is in reconstructing trajectories for a particular subset of molecular species, the $JSD^{(I_T)}$ for the subset can be used to assess the quality of the reconstruction relative to the other combinations of the measured species. This knowledge can help to refine the measurements, e.g., use smaller time intervals (see below) when the $JSD^{(I_T)}$ value for the subset of interest is not the minimum in the class. Thus, the $JSD^{(I_T)}$ value provides a valuable metric to quantitatively assess the quality of a trajectory reconstruction in experimental flow cytometry datasets where the correct trajectory is unknown. Using our *in silico* data we found that the errors become larger as the time interval is increased (Fig. S13A). This is expected as the slow variables start changing appreciably with the increasing magnitude of the time interval resulting in large values of the parameter $\xi_I (\gg 1)$. However, the precise answer to the question, if there exists an optimal time interval for cytometry measurements for good trajectory reconstructions needs further investigation (see Discussion). Lastly, we compared our method against a well-known scheme(8) that minimizes the total Euclidean distance between the measured variables at time $t_1$ and $t_2$ (Fig S13B-C). Our investigations produced mixed results. For a two species sub-module our method faired better (Fig S13B) whereas for a three species sub-module the quality of reconstruction between the two methods was comparable (Fig S13C). In few cases we found that the Euclidean method produced slightly better reconstructions.

## 3. Discussion
We have developed a framework for matching single cells across time using time-stamped data from flow and mass cytometry experiments. Our approach, based on posing the problem in terms of new variables that remain unchanged or vary slowly with time, is radically different from the existing methods (8, 11) employed for solving matching problems. Specifically, unlike other pair-matching algorithms(11), our approach does not require any assumption regarding the underlying reaction kinetics and estimation of model parameters. The use of slow variables and invariants reduces the value of parameter $\xi$ making the matching straightforward (when $\xi <1$) or more amenable to existing computational methods. The application of the framework for *in silico* signaling networks, live cell imaging data, and, synthetic flow cytometry data showed excellent to reasonable agreement of the reconstructed trajectories with the correct kinetics. We constructed two new variables from the measured species abundances that served as slow variables or invariants in a wide variety of signaling kinetics involving cell-to-cell variations of species abundances, nonlinear reaction propensities, and, intrinsic noise fluctuations. One of the new variables ($I_T$) is the total abundance of a molecular species in a single cell. Early time signaling events usually involve chemical modifications of signaling proteins (e.g., phosphorylation) that do not lead to any change in the total content of the proteins in a single cell. However, the total abundance of a protein at any time can vary from cell-to-cell over a wide range usually described by a log-normal distribution (25, 26). Thus, the total protein content provides a unique tag that can be used pair a single cell at any time with a "sister" cell at a later time. In flow and mass cytometry measurements, it is possible to assay the total content of certain signaling proteins (e.g., Erk), which can be used to reconstruct single cell signaling trajectories.

The other new variable ($I_M$) was constructed by scaling the measured single cell abundances by the inverse of the square root of the covariance matrix for species abundances. The reconstruction using $I_M$ worked better when the signaling kinetics in a time interval was effectively described by a closed system of first order reactions. Since $I_M$ behaves as a slow variable for a kinetics described by first order kinetics with small stochastic fluctuations and weak external fluxes, it can be used for addressing matching problems in other contexts such as tagged particles in fluid flows (8, 11). We recently developed a method to estimate reaction rates in a system of first order reactions designed to describe mass cytometry snapshot data in a time interval(27). The method also determines how well the system of first order reactions can describe the snapshot data. In case of a good agreement, we expect $I_M$ to behave as a slow variable and generate good to reasonable reconstructions. However, assessing if $I_M$ behaved as a slow variable for a cytometry data alone is not sufficient to infer the kinetics that can be described a set of first order reactions. The precise relation between $I_M$ being a slow variable and the underlying reaction network will require further investigation.

Our investigations of signaling kinetics in non-ideal cases showed that in several cases (Fig. S5) the reconstructed trajectories using $I_T$ produced better agreement to the correct kinetics as opposed to $I_M$. In these cases, the use of $I_T$ or $I_M$ reduced the value of the parameter $\xi$, however, $\xi^{(I)}$ still remained greater than 1, i.e., $\xi^{(I)}>1$. Thus, the cost function E (Eq. (10)) played an important role in matching the cells. The differences in quality of matching between the two slow variables point to the fact that minimizing E was better suited for $I_T$ compared to $I_M$ in these examples.

The application of our framework to synthetic flow cytometry data showed that $JSD^{(I_T)}$ values, evaluated for different subsets of measured species abundances can be used to determine groups of species where the reconstruction will produce less errors. The quality of the reconstruction improved as the size of the time intervals between successive measurements was decreased raising an important question about the existence of an optimal time interval. Similar questions in the context of selecting time points for measuring gene expressions have been dealt with machine learning methods where specific cost functions were optimized using finely spaced time measurements in a smaller gene subset(28). Such approaches might turn out to be helpful for generating trajectory reconstructions using cytometry data where live-cell imaging measurement of a smaller subset of proteins can be utilized to select time points in cytometry experiments measuring a large number of proteins simultaneously. Another question related to the above issue is if all the cells need to the coherently stimulated in cytometry experiments at time t=0 for good reconstructions. When the time difference in triggering for different batches of cell populations is long enough to produce large values in the parameter $\xi^{(I)}$, the reconstructions are likely to contain large errors. We also noticed deterioration (Fig. S13A) in the reconstruction quality as the kinetics approached the bistable behavior where the Ras activation changed by a large amount in a very short time interval. The framework is likely to be error prone when such abrupt changes occur in the kinetics.

In the recent years, a host of methods have been developed to construct single cell development trajectories using snapshot data (e.g., WANDERLUST(29), SCUBA(30)) where, unlike the cases dealt with here the single cells are not ordered temporally. These methods assign a "pseudo time" to the data and then optimize *ad hoc* cost functions to create single cell trajectories. It is unclear whether those cost functions will render any benefit to the matching problems considered here. For example, one of the cost functions that minimized the cosine distance between single cells (29) will not be able to correctly reconstruct signaling trajectories in a simple example where the kinetics are described by first order reactions.

The main difficulty in applying the framework developed here is to identify appropriate invariants or slow variables in a general situation. Singer et al.(31) used non-linear independent component analysis for constructing slow variables by analyzing stochastic kinetics of dynamical systems in a short time window. This method determined slow variables that were functions of linear combinations of the observables. When the underlying signaling reactions are known, this method can help find slow modes in the system by simulating the *in silico* network in short time durations, and, these slow variables can then be utilized to reconstruct single cell trajectories for cytometry data using our framework. However, the applicability of the method when the slow modes are complicated non-linear functions of the measured variables or when only a subset of involved dynamical variables is measured is unclear. In statistical physics(14, 15), conservation laws (e.g., conservation of energy, momentum) or symmetry principles help identify such slow modes. Projection operator methods by Zwanzig and Mori (32) provide a formal way to construct variables with slower time scales for a known microscopic dynamics. However, this method requires knowledge of model parameter values (e.g., reaction rates), and, the calculations for constructing slow modes could become intractable for a complex system composed of non-linear interactions such as cell signaling kinetics. A computation intensive step in our framework is to determine specific combinations of species abundances that are associated with low $JSD^{(I)}$ values. Mass-cytomety experiments can measure over 40 different proteins and the number of possible subsets in such large dimensions can be prohibitively large. When the signaling reactions are well characterized, selecting a group of species that are connected by mass balance in chemical modifications (e.g., enzymatic conversions or binding-unbinding reactions) could provide a way to identify a core species set with a slow mode. Adding new groups of species using smart sampling techniques(33) to expand this core set would be an exciting future endeavor.

**Materials and Methods:**

*1. Derivation of $I_M$ for the ideal kinetics.*

Eq. (1) can be solved analytically to calculate single cell abundances at any time t :

$$x_i^{(\alpha)}(t) = \sum_{j=1}^{n} \left[ e^{Mt} \right]_{ij} x_j^{(\alpha)}(0) \tag{14}$$

When the abundances follow Eq.(14), the average quantities in Eqs. (4)-(5) at the two time points ($t_1$ and $t_2$, $t_2 > t_1$) are related by,

$$\mu_i(t_2) = \sum_{j=1}^{n} \left[ e^{M(t_2-t_1)} \right]_{ij} \mu_j(t_1) \tag{15a}$$

$$J(t_2) = e^{M(t_2-t_1)} J(t_1) e^{M^T(t_2-t_1)} \tag{15b}$$

Notice that the elements of the M matrix cannot be uniquely determined from the above relations because there are $n^2$ unknown elements in the M matrix and Eqs.(15a-15b) provide $n+n(n+1)/2 < n^2$ relations between the unknown variables. Therefore, Eq. (14) cannot be used to evolve the abundances in a single cell at time point (at $t_1$) to a later time point ($t_2$), and then identify the correct cell from the measurements at $t_2$. Now, Eq. (15b) can be recast as,

$$e^{M(t_2-t_1)} = [J(t_2)]^{1/2} Q [J(t_1)]^{-1/2}$$

(16)

where, $Q$ is any orthogonal matrix, i.e., $QQ^T=I$, $I$ is the identity matrix. This equation contains the clue that if the abundances are scaled appropriately, the time evolution given by Eq. (1) can be described by a rotation or a reflection. We found that if the species abundances at any time are scaled by the inverse of the square root of the covariance matrix (Eq.(4)), and these scaled abundances are related across time points by orthogonal transformations.

Using Eq. (16) we can write Eq. (14) as,

$$x_i^{(\alpha)}(t_2) = \sum_{j=1}^{n} \left[ [J(t_2)]^{1/2} Q [J(t_1)]^{-1/2} \right]_{ij} x_j^{(\alpha)}(t_1)$$

(17)

which implies that the scaled variables in Eq. (3) at $t_1$ and $t_2$ are related by,

$$\tilde{x}_i^{(\alpha)}(t_2) = \sum_j Q_{ij} \tilde{x}_j^{(\alpha)}(t_1)$$

(18)

Since rotation and reflection preserves the magnitude of a vector, the magnitude ($I^{(\alpha)}_M$) of the vector $\overrightarrow{\tilde{x}^{(\alpha)}} \equiv \{\tilde{x}_1^{(\alpha)}, \tilde{x}_2^{(\alpha)}, \cdots, \tilde{x}_n^{(\alpha)}\}$ remains unchanged throughout the kinetics from its value at t=0.

*2. Simulation of the in silico networks.*

The mass action deterministic kinetics and the stochastic kinetics for the reactions for the system of first order reactions and the Ras activation network were simulated using the software package BIONETGEN(34). The initial species abundances were drawn from a multivariate normal distribution by specifying the average abundances and the covariances. The initial conditions and the parameter values for the reaction networks are provided in the supplementary material (Tables S1-S2 and Figs. S1 and S6).

*3. Generation of the synthetic flow cytometry data.*

The kinetics of Ras activation network (Fig. S6) was simulated using BIONETGEN(34). Single cell abundances of six different species (RasGTP, RasGDP, SOS, RasGRP1, DAG, and, RasGAP) were measured at different times (t=0, 100s, 200s, 300s, and, 500s). Different batches of 2000 single cells were used for measurements at two successive time points.

*4. Minimization of the cost function E.*

The minimization of E involves finding a bipartite graph where a pair of vertices in two subsets (single cells {α} at $t_1$ and single cells {β} at $t_2$) connected with a cost $(I^{(\alpha)}-I^{(\beta)})^2$ minimizes the total cost E. The graph matching algorithms computes the optimization in time $O(|E|\sqrt{V}) \approx O(n^2)$ (35). However, in our case we can use sorting to compute this in $O(n\ln(n))$ time. This is acheived by changing the cost function for a pairing #α, #β to $(I^{(\alpha)}-I^{(\beta)})^2 + \epsilon(t_2-t_1)$, where, $\epsilon \rightarrow 0$. Note, this is the Euclidean distance between the cells in the I-t plane, thus, minimizing the cost function E amounts to joining these points (or single cells) in the I-t plane by non-intersecting lines. Since, the cells in {α} (or {β}) have the same t co-ordinate t=$t_1$ (or $t_2$), the configuration with the non-intersecting lines keeps the same ascending (or descending) order in *I* for the cells in {α} and {β}. Therefore, first, we sorted the {α} and the {β} cells in arrays where cells were arranged in ascending order of *I*, and, then in the sorted arrays, we connected cells with the same array index. A pseudo-code is provided in the supplementary material. The Mathematica codes are available at http://planetx.nationwidechildrens.org/~jayajit/pair-matching.

*5. Calculation of JSD values.*

Calculation of $\text{JSD}^{(I_T)}$: For a subset of *m* molecular species ($m \leq n$) we calculated the sets $\{I_T^{(\alpha)}(t_1)\}$ and $\{I_T^{(\beta)}(t_2)\}$ for N=3000 cells. We then constructed the probability distributions $P(I_T, t_1)$ and $P(I_T, t_2)$ by binning the above sets and normalizing the histograms. The bin width ($\Delta I_T$) was chosen to be the cardinality k of the subset. We then calculated the Kullback Leibler divergence ($d_{KL}$) using the distributions. $d_{KL}$ is given by an integral here as $I_T$ is a continuous variable. We approximated the integral by a sum over the histograms calculated above, i.e.,

$$\int_{I_T^{min}}^{I_T^{max}} P(I_T) \ln\left[\frac{P(I_T)}{M(I_T)}\right] dI_T \approx \Delta I_T \sum_{p=I_T^{min}}^{I_T^{max}} P((I_T)_p) \ln\left[\frac{P((I_T)_p)}{M((I_T)_p)}\right]$$

where $\Delta I_T$=k and $M(I_T)=[P(I_T(t_1))+P(I_T(t_2))]/2$.

Calculation of $\text{JSD}^{(I_M)}$: Done in a similar way as $I_T$. The bin width ($\Delta I_M$) was chosen as follows. Since,

$$\Delta I_M^{(\alpha)} = \frac{\sum_{i,j}\left(\Delta x_i^{(\alpha)} J_{ij}^{-1} x_j^{(\alpha)} + x_i^{(\alpha)} J_{ij}^{-1} \Delta x_j^{(\alpha)}\right)}{2|\vec{\tilde{x}}^{(\alpha)}|} = \frac{\sum_{i,j}\left(\Delta x_i^{(\alpha)} J_{ij}^{-1} x_j^{(\alpha)}\right)}{|\vec{\tilde{x}}^{(\alpha)}|}$$

Using

$$I_M^{(\alpha)} = \left|\vec{\tilde{x}}^{(\alpha)}\right| = \sqrt{\sum_{i,j} x_i^{(\alpha)} J_{ij}^{-1} x_j^{(\alpha)}}, \quad \Delta x^{(\alpha)} = 1 \text{ for all species and all the cells}$$

$$\Delta I_M^{(\alpha)} = \frac{\sum_{i,j} J_{ij}^{-1} x_j^{(\alpha)}}{\left|\vec{\tilde{x}}^{(\alpha)}\right|}$$

We evaluated $\Delta I_M^\alpha$ for all the cells at time $t_1$ and $t_2$ and set $\Delta I_M = \min(\Delta I_M^{(\alpha)})$. A pseudo-code is provided in the supplementary material. The Mathematica codes are available at http://planetx.nationwidechildrens.org/~jayajit/pair-matching.

**Figure legends**

**Fig. 1: Matching single cells across time using invariants and slow variables. A)** Schematic depiction of time-stamped cytometry data showing single cells against copy numbers (denoted by $x_1$ and $x_2$) of two signaling species at three different time points $t_1 < t_2 < t_3$. When the average distance ($l_0$) between the cells in the $x_1$-$x_2$ plane is smaller than the average distance traveled ($\Delta l$) in a time interval (e.g., $t_2$-$t_1$) by the cells in the same plane or $\xi = \Delta l / l_0 \gg 1$, matching the cells across time is non-trivial due to multiple possibilities. Whereas when $\xi < 1$, connecting the cells could be as straightforward as finding the nearest neighbor. The arrows show a correct trajectory. **B)** We cast the matching problem in (A) in the manifold for a new variable (I) constructed from the measured variables (e.g., $x_1$, $x_2$). $I$ satisfies two conditions: (i) $I$ does not change (invariant) or changes substantially slowly (slow variable) in individual cells compared to the original variables or $\Delta l^{(I)} \ll \Delta l$, and, (ii) $I$ varies between single cells at any time point. In this situation $\Delta l^{(I)}/l_0^{(I)} = \xi^{(I)} \ll 1$ and connecting the cells across time becomes straightforward. When $I$ is an invariant, $\Delta l^{(I)} = 0$ and the pairing of cells across time is exact. **C)** In many cases, $I$ will be a slow variable resulting in $\xi^{(I)} > 1$. However, even in such cases we find, $\xi^{(I)} < \xi$, and a cost function is employed to approximately pair single cells across time.

**Fig. 2: Determination of slow variables or invariants: A)** Signaling networks are composed of a large number of species, however, only few of them can be assayed. In the schematic network of 6 different species, only four species (shown with yellow boxes) $x_1$, $x_2$, $x_4$, and, $x_6$ can be measured in experiments. Cytometry experiments measure single cell abundances of these species (also denoted as $\{x_i\}$ here). **B)** $2^4$-4-1=11 different subsets of measured species abundances can be constructed. Each of these subsets represents projection of the original data into a manifold spanned by the member species abundances in the subset. The subsets are further divided into classes for our analyses.

Each class contains subsets with the same cardinality. For example, class#3 contains all the three species subsets. **C)** For each subset we evaluate the change in the distribution of the species abundances in the cell population using JSD (Eq. (7)). A large change in the distribution of species abundances indicating faster kinetics will result in a larger value of JSD. Whereas, if $I$ behaves as a slow variable or an invariant, the distribution of $I$ in the cell population will go through a small change or no change, respectively. The change in $I$ is quantified by $JSD^{(I)}$. When $JSD^{(I)}$ is smaller than the slowest species in the subset, we denote $I$ as a slow variable for that subset.

**Fig. 3: Pairing single cells for the ideal kinetics. A)** Heat map for $JSD^{(I_M)}$. JSD was calculated for 3000 single cells (see Materials and Methods section for details) at time $t_1=0$ and $t_2=7$mins. Other parameters related to the signaling kinetics are shown in the web supplement. **B)** Shows the minimum values of $JSD^{(I_M)}$ for each class (grey points). JSD values associated with the fastest (shown in orange) and the slowest species (shown in blue) in the subsets corresponding to the minimum $JSD^{(I_M)}$ are compared with minimum $JSD^{(I_M)}$. **C)** Shows the quality of matching when $I_M$ was used for matching the cells. The ratio in the average error $\overline{\chi}$ in the matching using $I_M$ with that for random matching is shown for each of the subsets. $\overline{\chi}<1$ indicates smaller error compared to the random pairing. **D)** Error in the autocorrelation function ($\Delta$) for the subsets corresponding to the minimum values of $JSD^{(I_M)}$. $\Delta^{random}$ denotes the error in the autocorrelation function for random matching.

**Fig. 4: Pairing single cells for non-ideal kinetics. A)** Shows the minimum values of $JSD^{(I_T)}$ for each class (black points) for the deterministic Ras activation kinetics model. We used 3000 single cells across time points $t_1=100$s and $t_2=400$s where the Ras activation displays bistability. $\text{Min}(JSD^{(I_T)})$ values were compared with the JSD values associated with the fastest (shown in orange) and the slowest species (shown in blue) in the subsets corresponding to $\min(JSD^{(I_T)})$. $I_T$ behaves as an invariant or a slow variable for the subsets. **B)** Shows the quality of matching when $I_T$ was used for matching the cells for the subsets associated with $\min(JSD^{(I_T)})$ in (A). **C)** Error in the autocorrelation function ($\Delta$) when the cells were matched using $I_T$ for the stochastic Ras activation kinetics. The subsets used in the matching yielded minimum values of $JSD^{(I_T)}$ for each class. Note that for exact matching ($\Delta=0$) $I_T$ behaved as an invariant.

**Fig. 5. Trajectory reconstruction using live-cell imaging data. A)** Schematic diagram for gene activation. **B)** Reconstruction of a typical kinetics trajectory for CFP and YFP tagged dual reporter for the gene DCS2 in yeast diploid cells. The re-constructed trajectory using $I_T$ (shown in orange) is compared with the true trajectory (shown in blue).

**C)** Distribution of the quality of alignment χ using $I_T$ is compared to the quality of alignment using random pairing for 159 single cell trajectories. The reconstructions were carried out between two successive time measurements for 63 time intervals (e.g., 0min to 2.5min, 2.5min to 5min, and so on). **D)** Shows the ratio of the errors (Eq. 13) in the auto-correlation function for trajectories reconstructed using $I_T$ with that for random pairing for the same time intervals in (C). $\Delta A/\Delta A_{random} < 1$ for most of the time points indicating better matching using $I_T$ compared to the random pairing.

**Fig. 6. Reconstruction using synthetic flow cytometry data. A)** Shows the variation of the ratio of the average relative error $\bar{\chi}_{Rel}^{JSD_{min}} / \bar{\chi}_{Rel}^{JSD_{max}}$ with the class k. The data were measured at t=0 and t=100s. $\bar{\chi}_{Rel}^{JSD_{min}}$ and $\bar{\chi}_{Rel}^{JSD_{max}}$ for a class k were calculated for the species subsets that generated the smallest and the largest values of $JSD^{(I_T)}$, respectively. The smaller than 1 values for the ratio indicate that the errors in the subset corresponding to $JSD_{min}$ were smaller on average than that for the subset with the largest JSD in the same class. The subsets the corresponded to $JSD_{min}$ are the following: {RasGDP, RasGAP} for k=2, {RasGDP, RasGTP, RasGAP} for k=3, {RasGDP, RasGTP, RasGAP, DAG} for k=4, {RasGDP, RasGTP, RasGAP, DAG, RasGRP1} for k=5, and, {RasGDP, RasGTP, RasGAP, DAG, RasGRP1, SOS} for k=6. The relative error $\chi_{Rel}^{\alpha\beta}$ between a cell# $\alpha$ at $t_1$ and its matching partner cell# $\beta$ (correct partner cell# $\beta'$) at $t_2$ is defined as,

$$\chi_{Rel}^{\alpha\beta} = \sqrt{\sum_{i=1}^{k}\left(x_i^{(\beta)} - x_i^{(\beta')}\right)^2} \bigg/ \left|x_i^{(\beta')}\right|$$

We used this definition instead of Eq. 11 to compare between the subsets containing abundances of very different values. The average value of $\bar{\chi}_{Rel}$ was calculated by taking the average of $\chi_{Rel}^{\alpha\beta}$ values for all the single cell pairs. **B)** Shows the ratio $\bar{\chi}_{Rel}^{JSD_{min}} / \bar{\chi}_{Rel}^{Rand}$ for the pairing for subsets associated with minimum JSD in (A). $\bar{\chi}_{Rel}^{Rand}$ is the average relative error when the matching was generated using random pairing of the cells. The ratio is <1 for all the cases indicting that the reconstruction is better than random reconstructions.

*Ethics Statement.* Not applicable.

*Data availability.* The computer codes used for the studies shown the article are available at the website, http://planetx.nationwidechildrens.org/~jayajit/pair-matching . The codes are written in Mathematica and in BIONETGEN (bionetgen.org). The live-cell imaging data reported in Fig. 4 in Ref. (17) are available on the web at the link, http://msb.embopress.org/content/9/1/704.short.

*Competing Interests.* We have no competing interests.

*Author's contributions.* SM and JD planned and designed research. SM and JD performed analytical and numerical calculations. SM, WS, LLL, and, JD, analyzed data. SM, WS, DS, and, JD contributed computational tools. SM, LLL, and, JD wrote the paper. All authors made significant contributions to the manuscript. All authors gave their final approval for publication.

*Acknowledgements.* JD and SM thank Helle Jensen, Suzanne Gaudet, Alper Yilmaz, Victoria Best, and, Anton Zilman for helpful discussions. SM and JD acknowledge the help from Anders Hansen in Erin O'Shea's lab for accessing the live imaging data. JD also acknowledges the support from the Quantitative Immunology Workshop at KITP where a part of the work was carried out.

*Funding statement.* This work was supported by the grant R56AI108880-01 from NIAID. SM was supported in part by a grant from the Department of Biotechnology (BTPR12422/MED/31/287/2014, valid November 2014 to 2017). We also acknowledge the computation time provided by the Ohio Supercomputer Center.

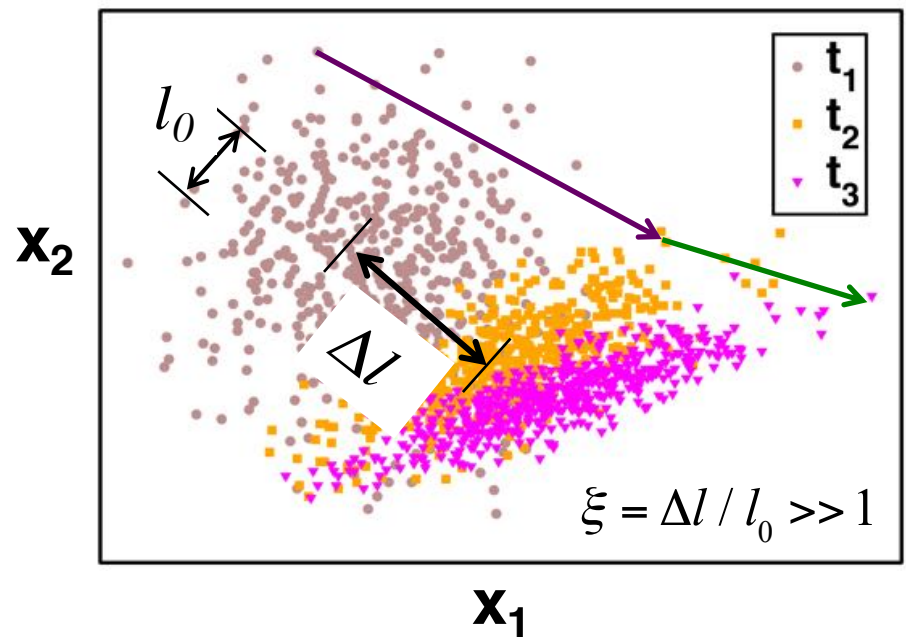
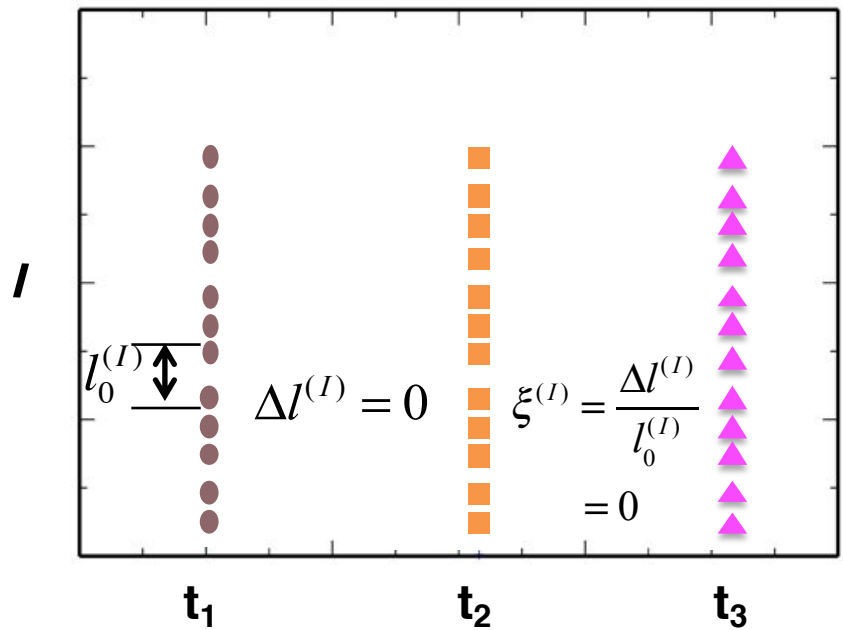
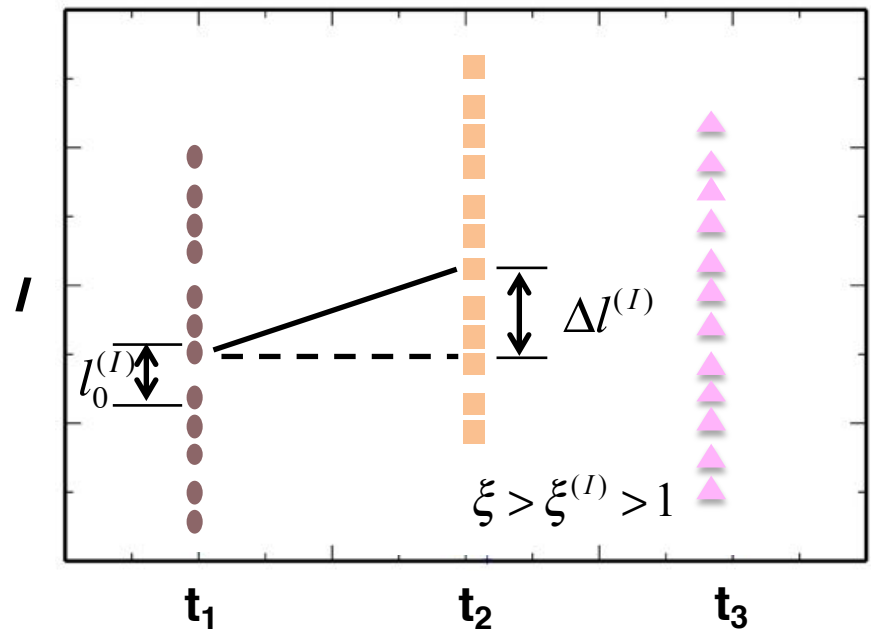

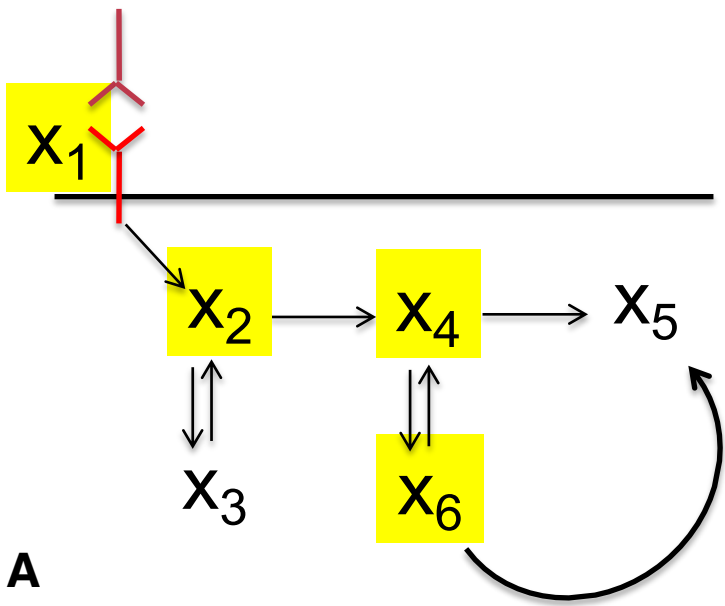
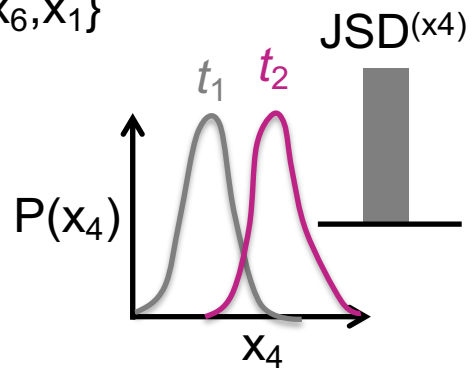
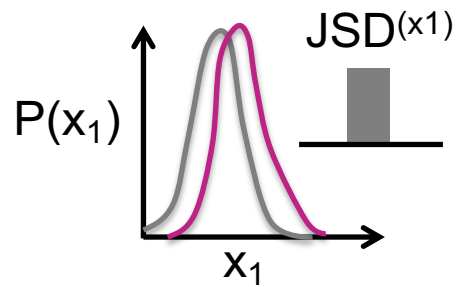
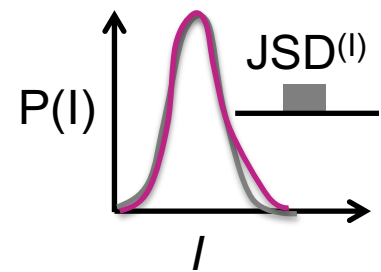

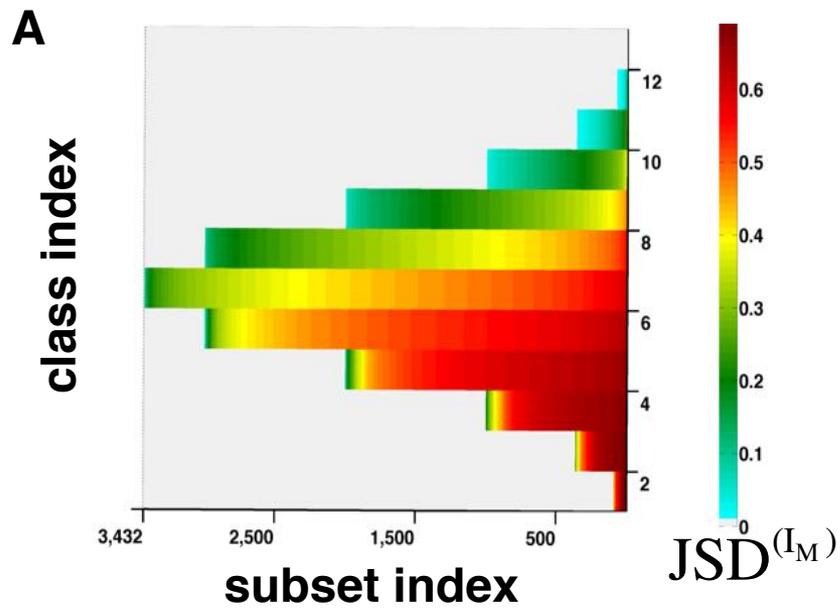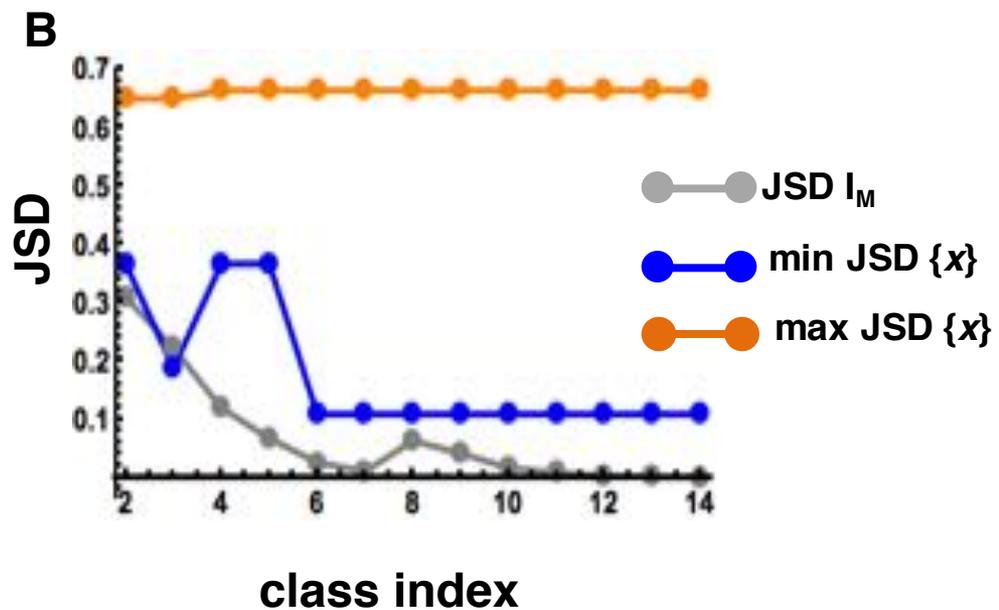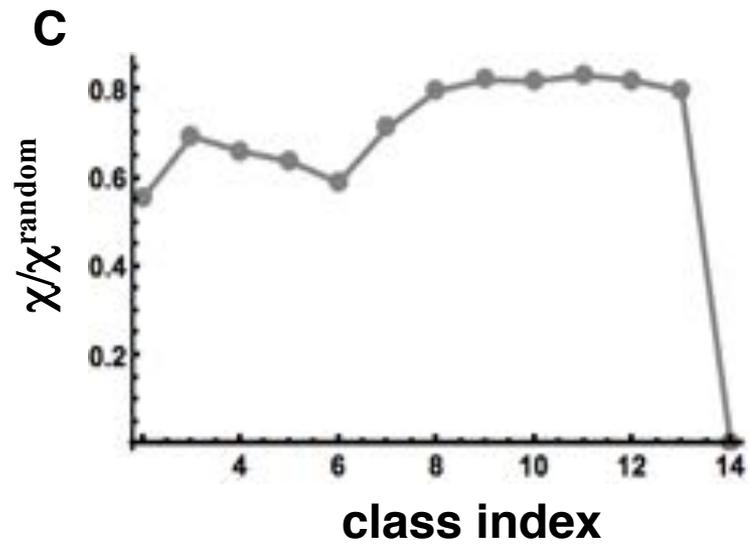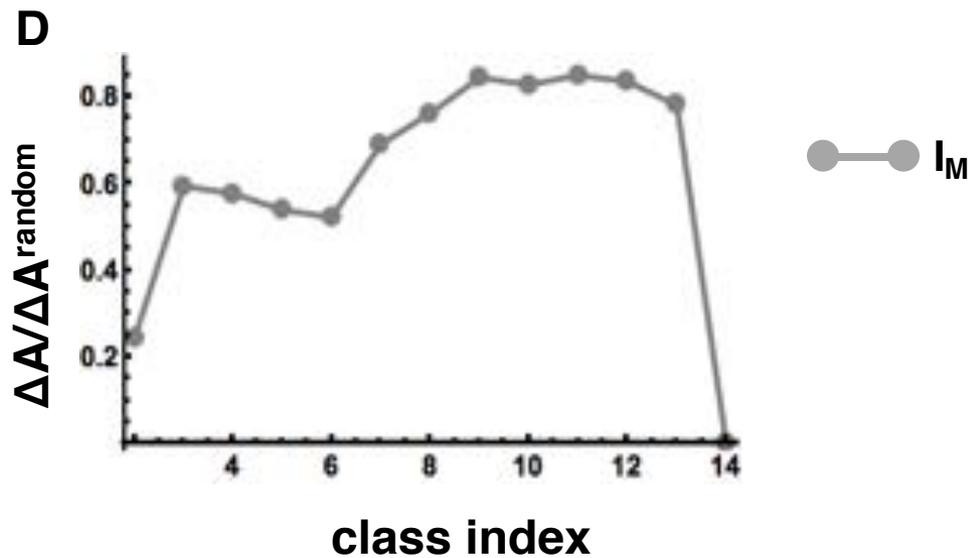

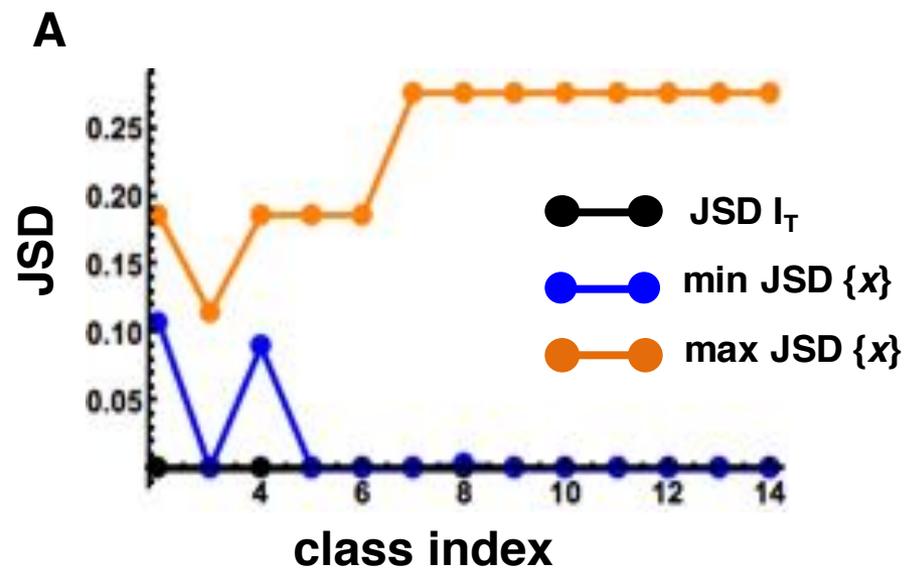
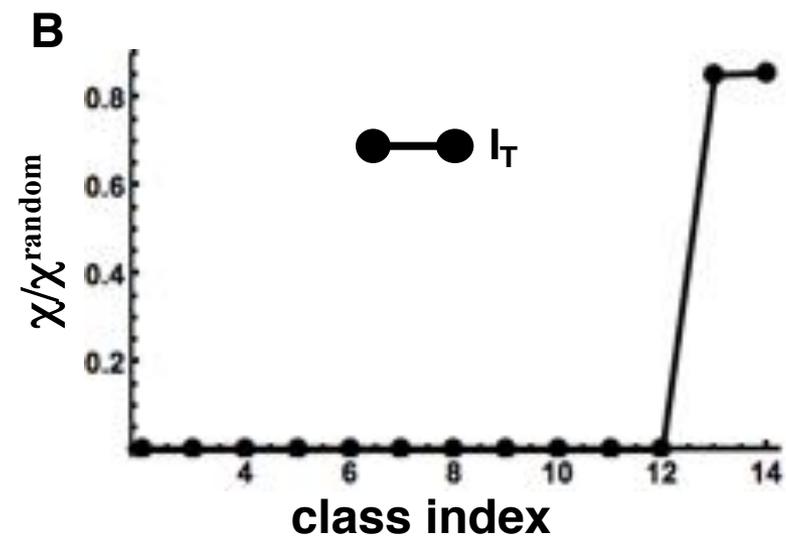
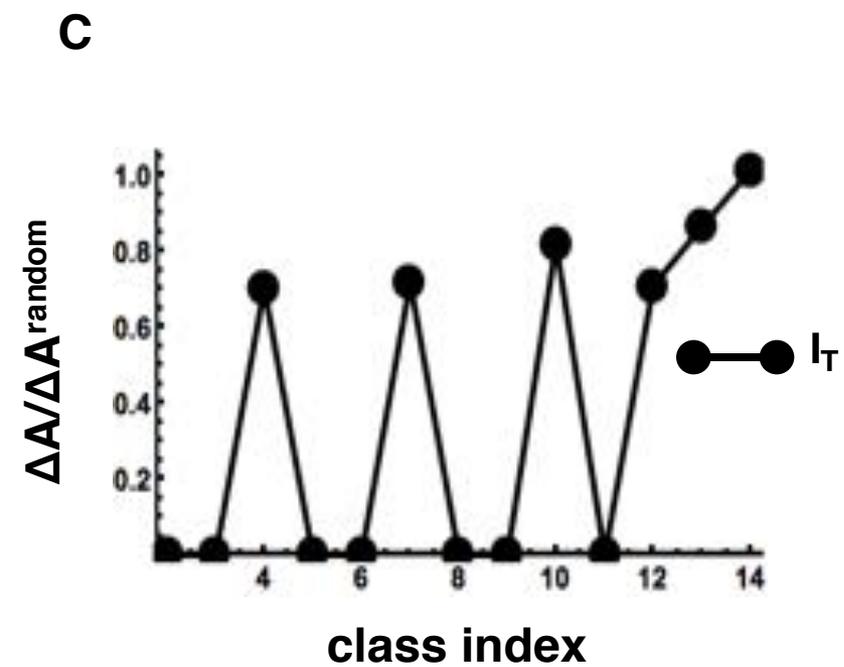

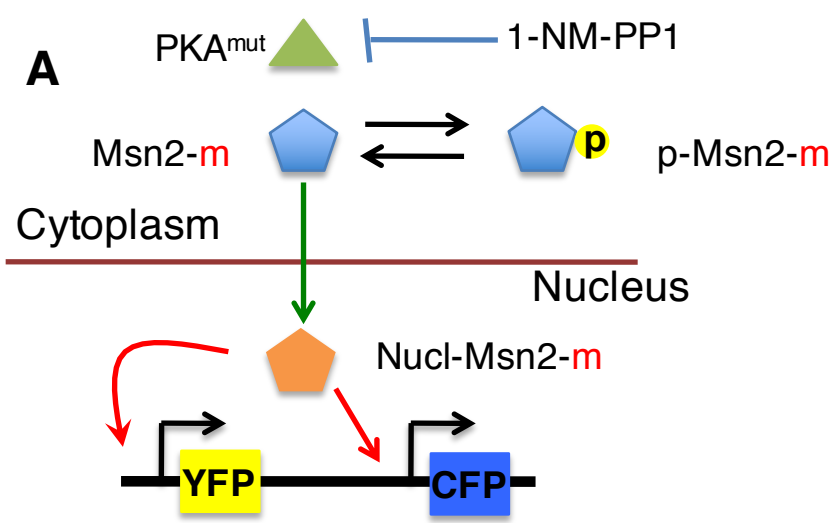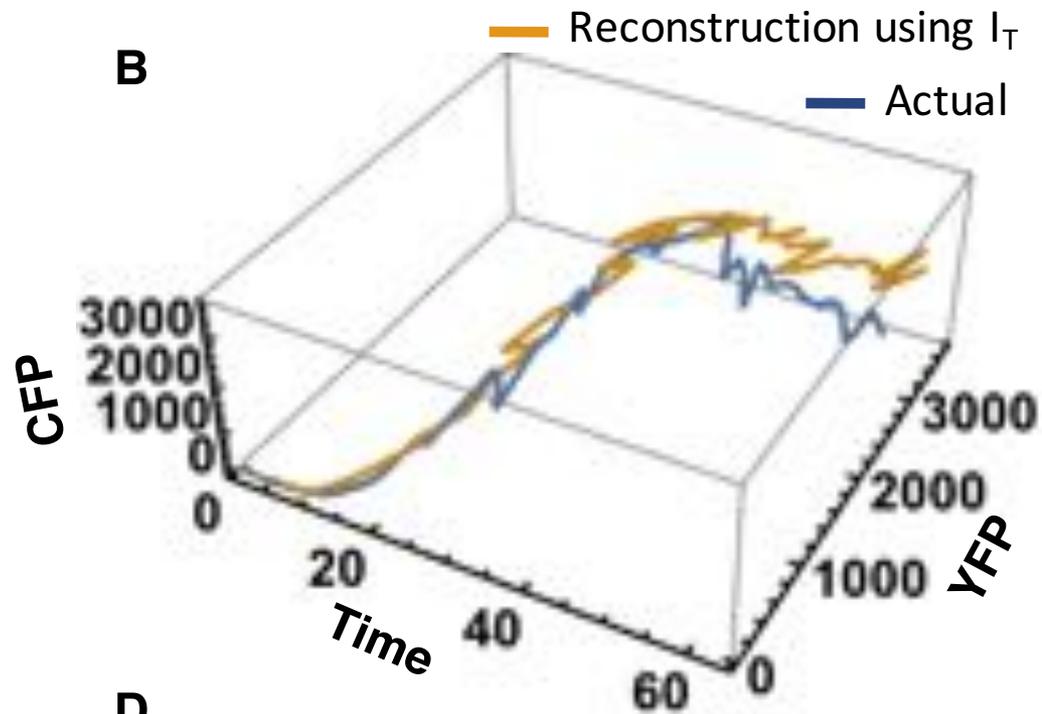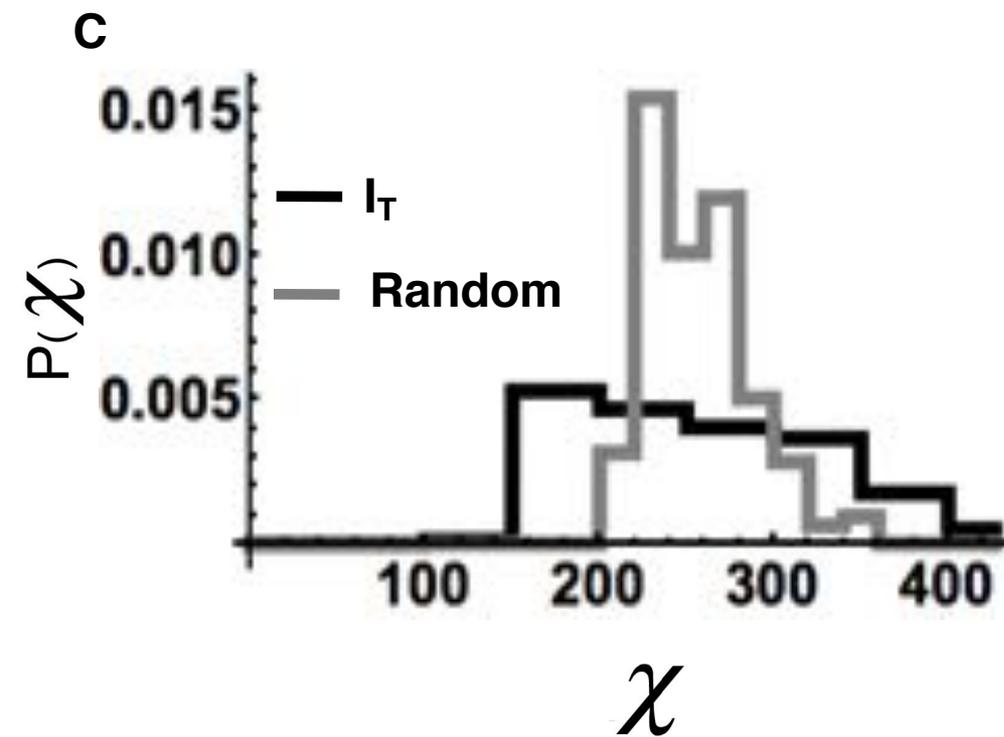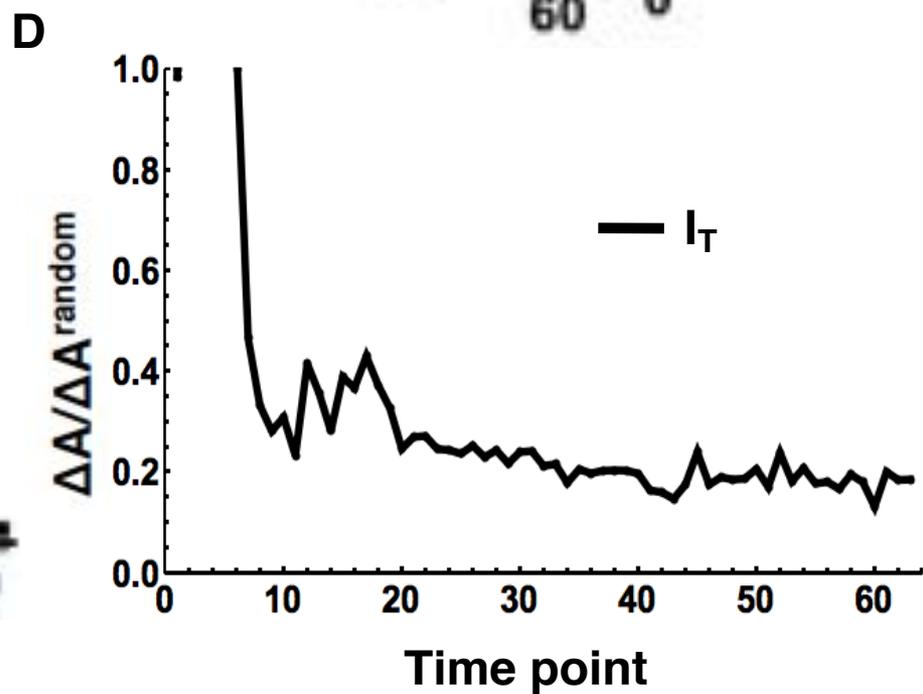

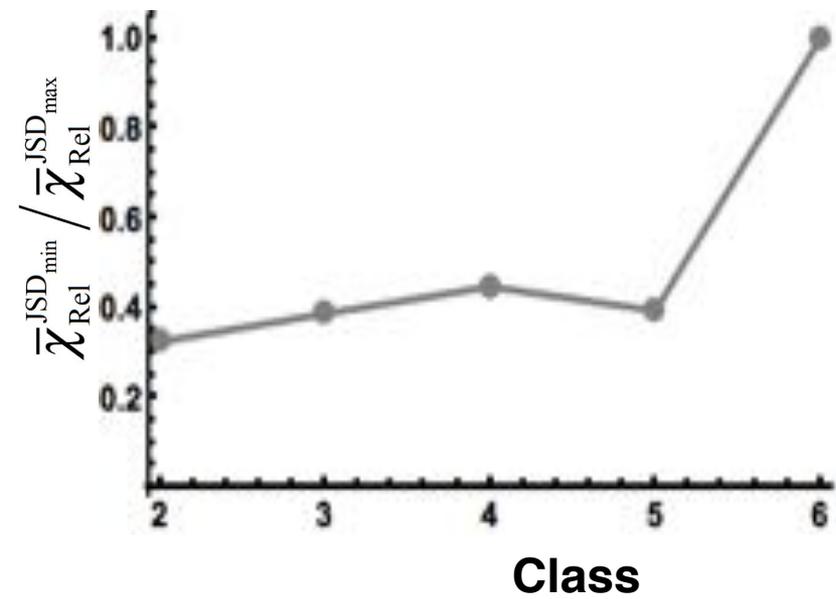 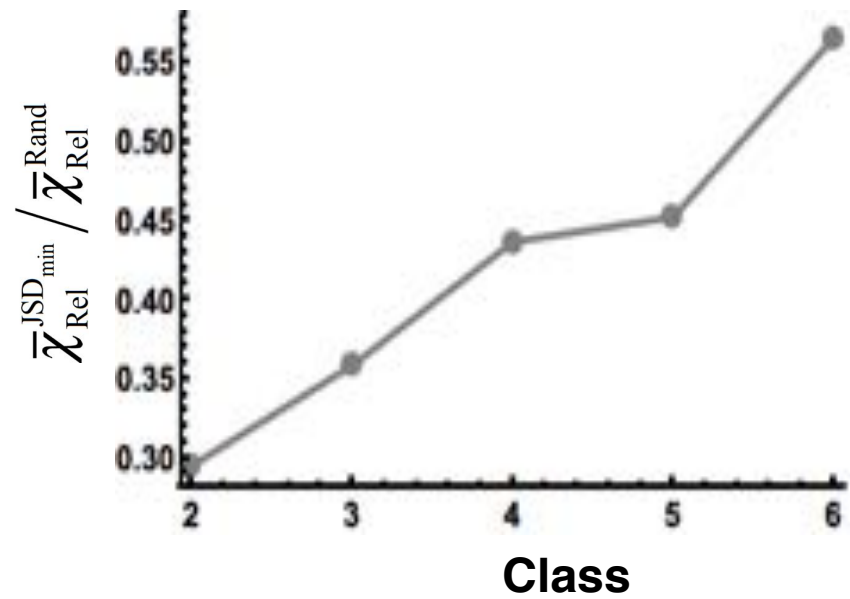

# Supplementary Material for "Connecting the dots across time: Reconstruction of single cell signaling trajectories using time-stamped data"

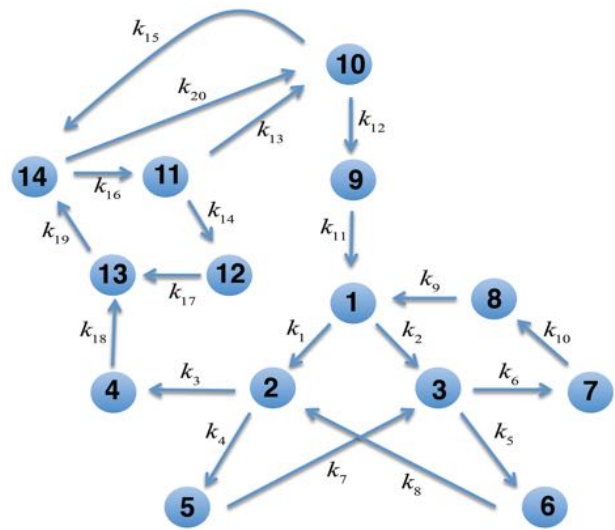

**Figure S1: Wiring diagram of the network of first order reactions.** The species are indicated by the integers and the rate constants shown as $\{k_a\}$. The values for the rate constants and the initial abundances are shown in Table S1.

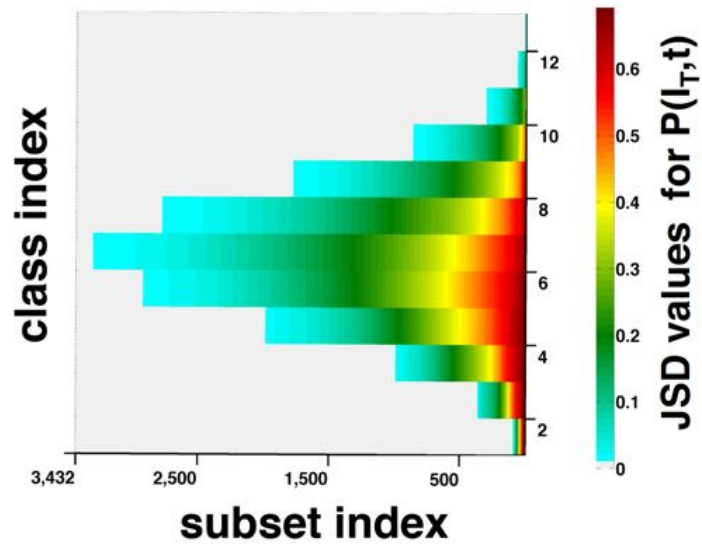

**Figure S2: Variation of $JSD^{(I_T)}$ with subsets in the linear network**. The calculation was carried out for the 16369 different subsets of signaling species involving 3000 single cells at a pair of time points ($t_1$=0, $t_2$=7min). The kinetics is described by first order reactions corresponding to the network in Fig. S1 and Table S1.

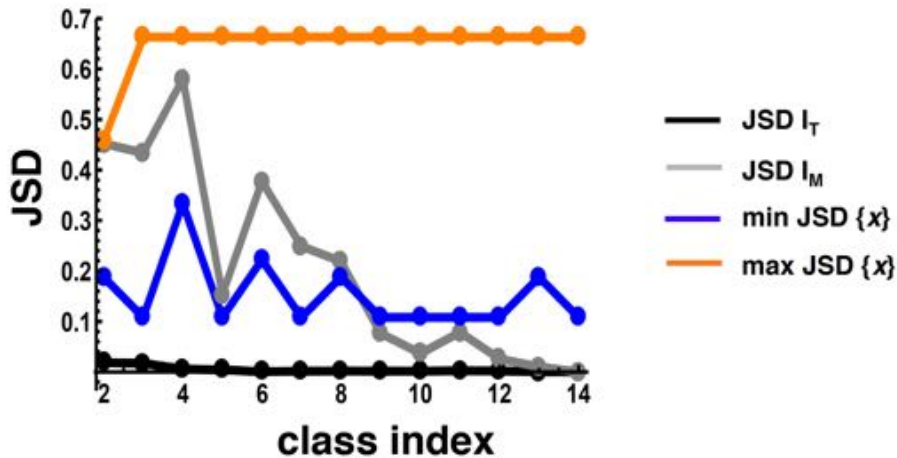

**Figure S3: $I_T$ behaves as a slow variable or an invariant in the linear kinetics.** Shows the minimum values of $JSD^{(I_T)}$ for each class (black points) for the kinetics with first order reactions. The parameter values are the same as in Fig. S2. JSD values associated with the fastest (shown in orange) and the slowest species (shown in blue) in the subsets corresponding to the minimum $JSD^{(I_T)}$ are compared with minimum $JSD^{(I_T)}$. The grey line shows the values of $JSD^{(I_M)}$ for the subsets that yielded the min($JSD^{(I_T)}$). For multiple subsets (e.g., the subsets corresponding to class#10, #12, #13), both $I_T$ and $I_M$ behave as slow variables.

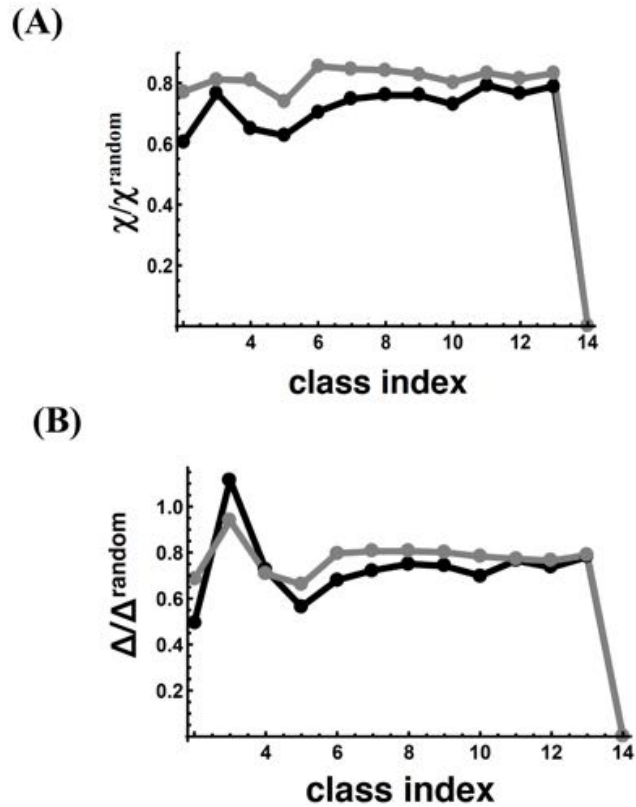

**Figure S4: Result of pairing using $I_T$ for the linear kinetics.** The reconstruction was carried out using $I_T$ for the subsets that produced min $JSD^{(I_T)}$ values (Fig. S3). **(A)** The ratio (black points) of the average error in the reconstruction with that for random pairing corresponding to subsets. The grey points show the error when the reconstruction was carried out for the same subsets using $I_M$ instead of $I_T$. (B) Shows error in the autocorrelation function for the same reconstructions in (A).

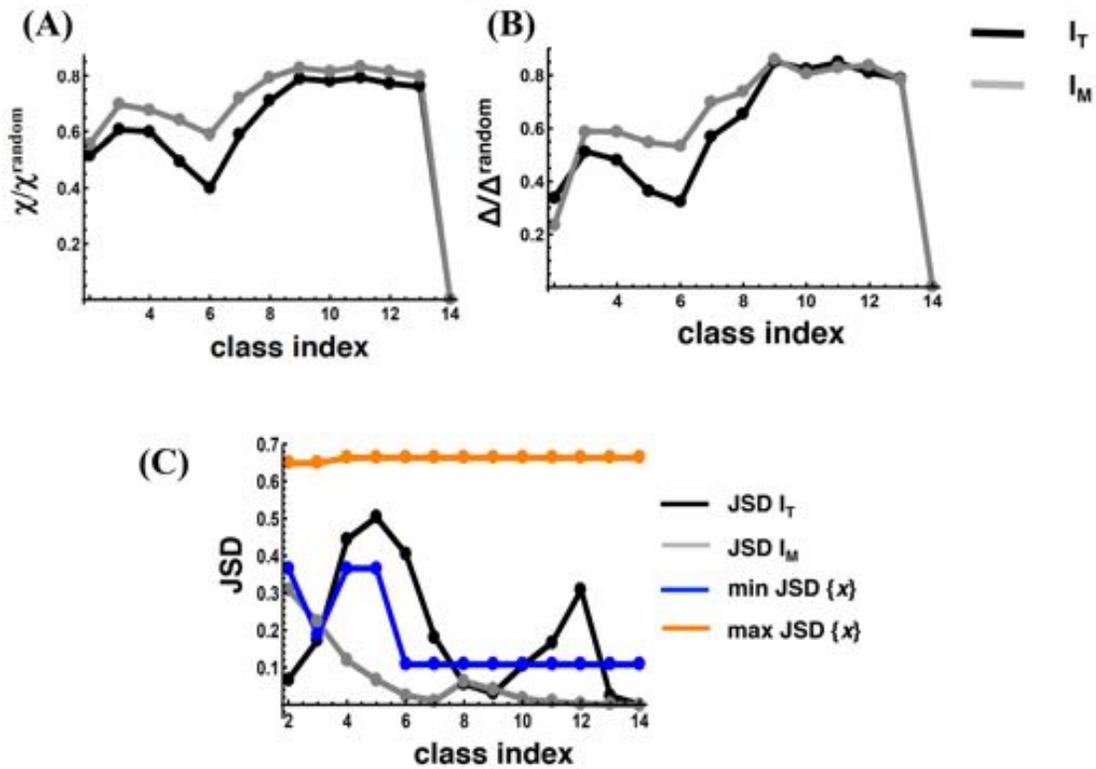

**Figure S5: Comparison of the reconstructions carried out using $I_M$ or $I_T$ for the linear network.** The reconstruction was carried out for the subsets that produced minimum $JSD^{(I_M)}$. The parameters are the same as in Fig 2. (A) Shows the error (grey points) when the reconstruction was performed using $I_M$. The black points show the errors when the single cells were paired using $I_T$ for the same subsets that yielded minimum $JSD^{(I_M)}$. (B) Shows error in the autocorrelation function for the reconstructions in (A). The pairings using $I_T$ show lower error even though the subsets produced minimum values of $JSD^{(I_M)}$. (C) Shows the $JSD^{(I_T)}$ values (black points) for the subsets that produced minimum values of $JSD^{(I_M)}$. In several cases (e.g., subset for class #4) $I_T$ has a faster kinetics than $I_M$.

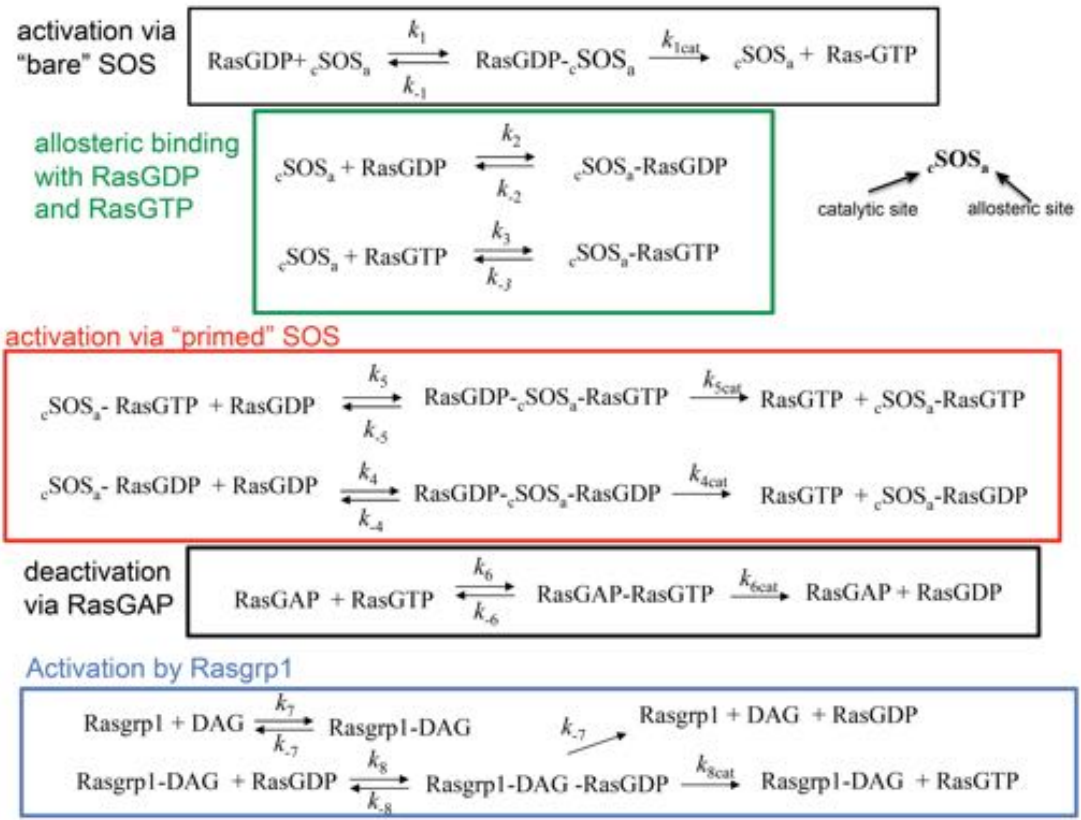

**Figure S6: The Ras activation network.** SOS and Rasgrp1 are the enzymes that convert the de-active form of Ras or RasGDP to the active form, RasGTP. SOS binds RasGDP at the catalytic site and an allosteric site. In addition, the allosteric site in SOS can bind to RasGTP. When the allosteric site of SOS is occupied by RasGTP or RasGDP, the catalytic rate for Ras activation is increased. This creates a positive feedback in the activation. Rasgrp, when bound to the membrane bound DAG, acts as an enzyme for Ras activation. Ras de-activation is carried out by the enyme RasGAP. The reactions and the rates are shown above. The rates and initial conditions are shown in Table S2.

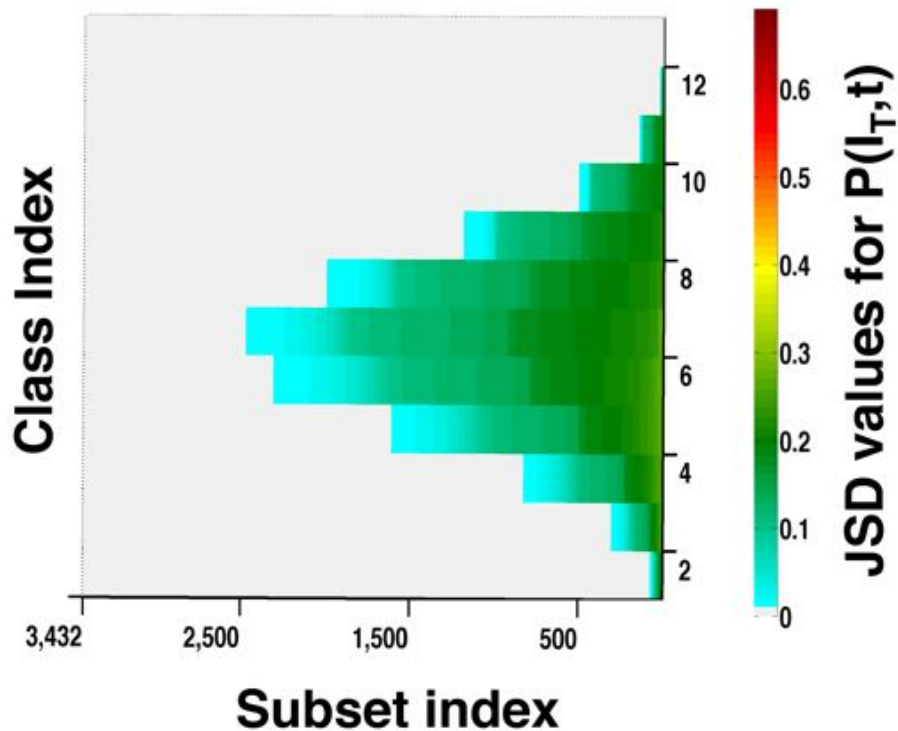

**Figure S7: Variation of $JSD^{(I_T)}$ for the deterministic Ras activation kinetics.** We used 3000 single cells across time points $t_1$=100s and $t_2$=400s where the Ras activation displays bistability. The kinetics was simulated using the reactions shown in Fig. S6 using the software package BIONETGEN.

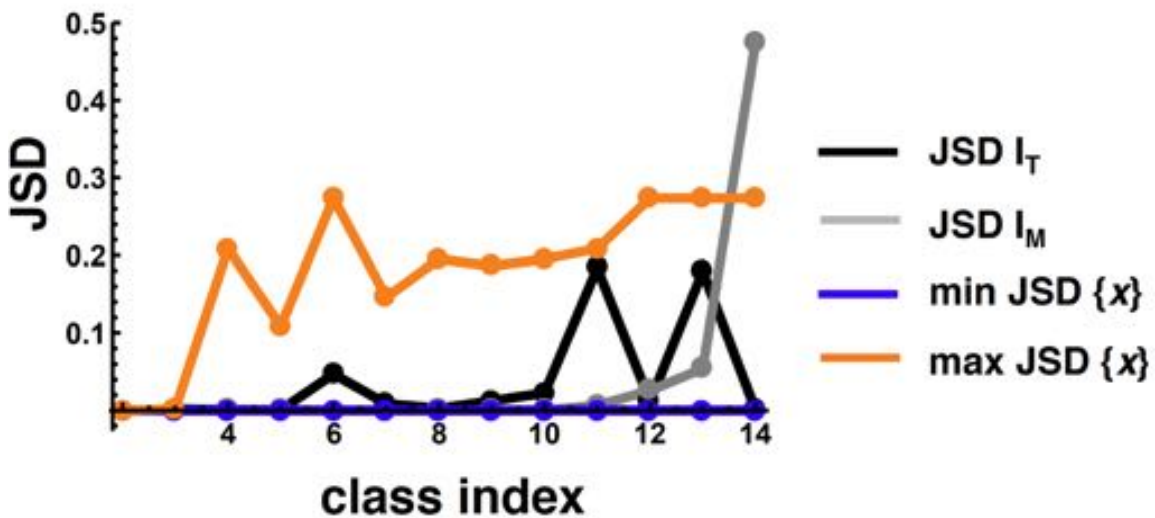

**Figure S8: $I_M$ behaves as a slow variable for the deterministic Ras activation kinetics.** Shows the minimum values of $JSD^{(I_M)}$ (grey points) for all the subsets. The JSD values

corresponding to the fastest (orange points) and the slowest species (blue points) are shown for comparison. Some of the species abundances reached values closer to the steady state and generate very small change in the time window. When present, these species corresponded to the slowest species. $JSD^{(I_M)}$ values were equal (very close to zero) to the JSD values for the slowest species for the subsets until class#10. Thus, $I_M$ behaved as slow variables in these subsets. $JSD^{(I_T)}$ values for the subsets generating minimum $JSD^{(I_M)}$ are shown in black.

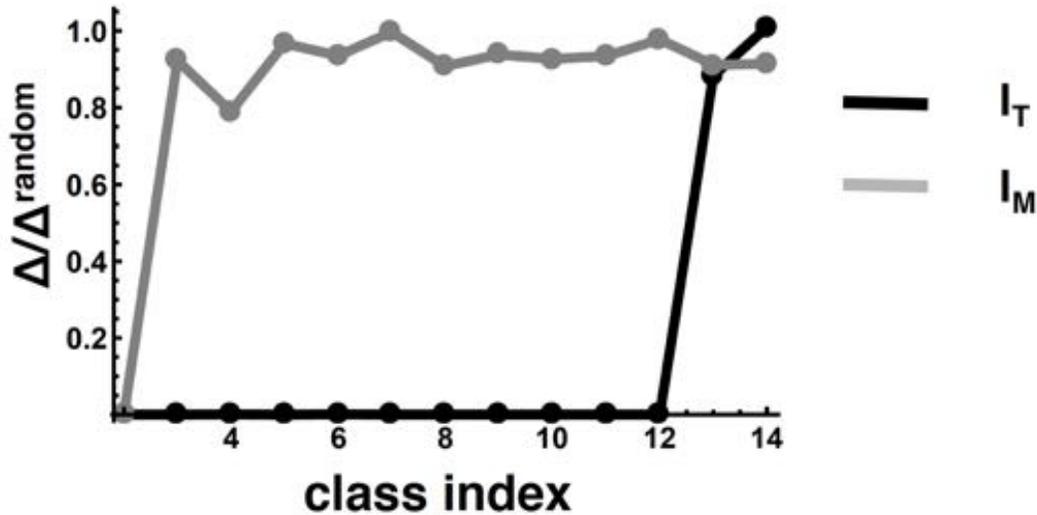

**Figure S9: Error in the autocorrelation for the pairings performed using $I_T$ and $I_M$ for the deterministic Ras activation kinetics.** Error in the autocorrelation function when the pairings were performed using $I_T$ and $I_M$. The pairings were carried out for the subsets that yielded minimum values for $JSD^{(I_T)}$ (black) or $JSD^{(I_M)}$ (grey). The subsets are shown with their class indices. Note, for each class the subsets used for $I_T$ or $I_M$ are different. The data compare the best level of pairings that can be carried out by $I_T$ or $I_M$. Except the subsets with 2, 13, and, 14 species (class #2, #13, and #14, respectively), pairing by $I_T$ performs substantially better than that by $I_M$.

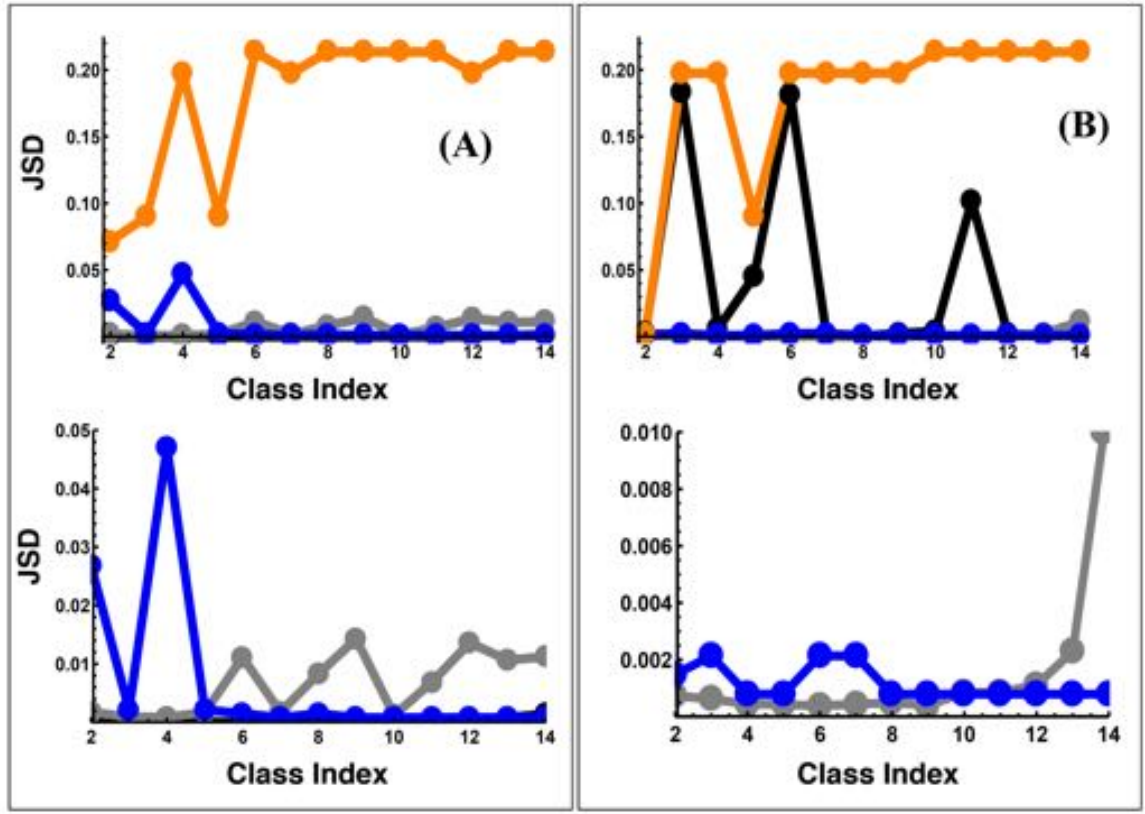

**Figure S10: $I_T$ and $I_M$ behaves as a slow variable for stochastic Ras activation kinetics.** Ras activation kinetics including intrinsic noise fluctuations was generated using the same reaction network used for investigating the deterministic kinetics. The simulations were carried out using the software package BIONETGEN. The parameters used for the kinetics are the same as that used for Fig. S8. **(A)** Shows the minimum values of $JSD^{(I_T)}$ (black points) for all the subsets. The values for $JSD^{(I_T)}$ are very close to zero and are covered by the grey or the blue symbols. The JSD values corresponding to the fastest (orange points) and the slowest species (blue points) are shown for comparison. The $JSD^{(I_M)}$ values corresponding to the subsets for min ($JSD^{(I_T)}$) are shown in grey. The separation between the points at the lower JSD values are shown in the bottom panel where the y axis is zoomed in between 0 to 0.5. **(B)** Shows the minimum values of $JSD^{(I_M)}$ (grey points) for all the subsets. The JSD values corresponding to the fastest (orange points) and the slowest species (blue points) are shown for comparison. The $JSD^{(I_T)}$ values corresponding to the subsets for min ($JSD^{(I_M)}$) are shown in black. The bottom panel shows the zoomed in version of the top graph (grey and blue points only) for the points at the lower JSD values.

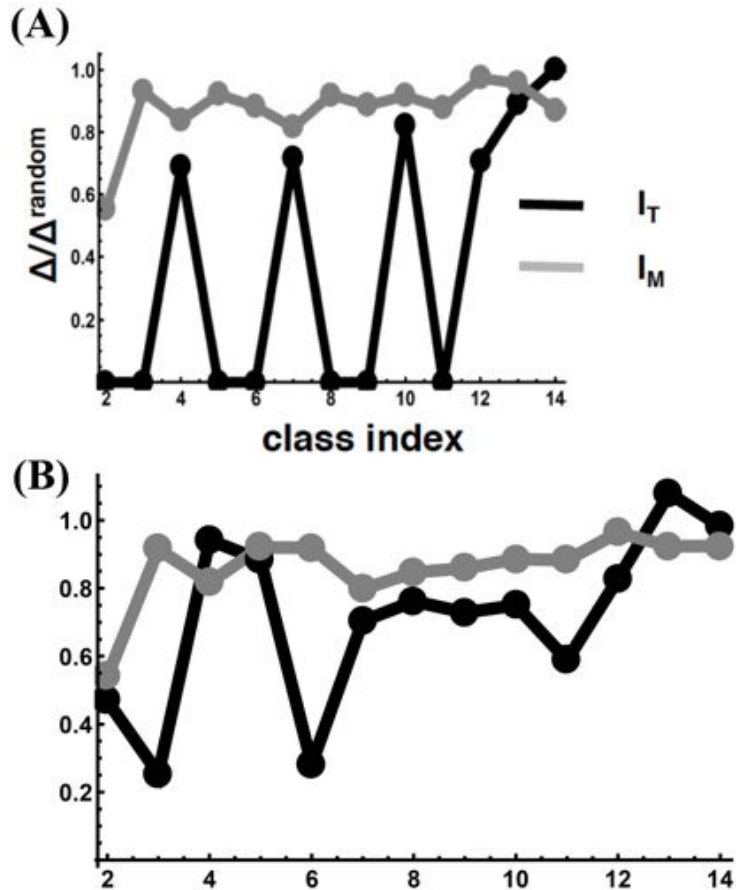

**Figure S11: Error in the autocorrelation function for the pairings performed using $I_T$ and $I_M$ for the stochastic Ras activation kinetics.** The pairings were carried out for 3000 single cells for stochastic Ras activation kinetics. **(A)** Pairing was done using $I_T$ for the subsets (indexed by the class numbers) that yielded minimum values of $JSD^{(I_T)}$. Error in the autocorrelation function (black points) for single cell-sister cell pairs. The grey points show the error when $I_M$ was used for pairing the single cells for the same subsets that yielded minimum values of $JSD^{(I_T)}$. **(B)** Pairing was done using $I_M$ for the subsets (indexed by the class numbers) that yielded minimum values of $JSD^{(I_M)}$. Error in the autocorrelation function (grey points) for single cell-sister cell pairs. The black points show the error when $I_T$ was used for pairing the single cells for the same subsets that yielded minimum values of $JSD^{(I_M)}$.

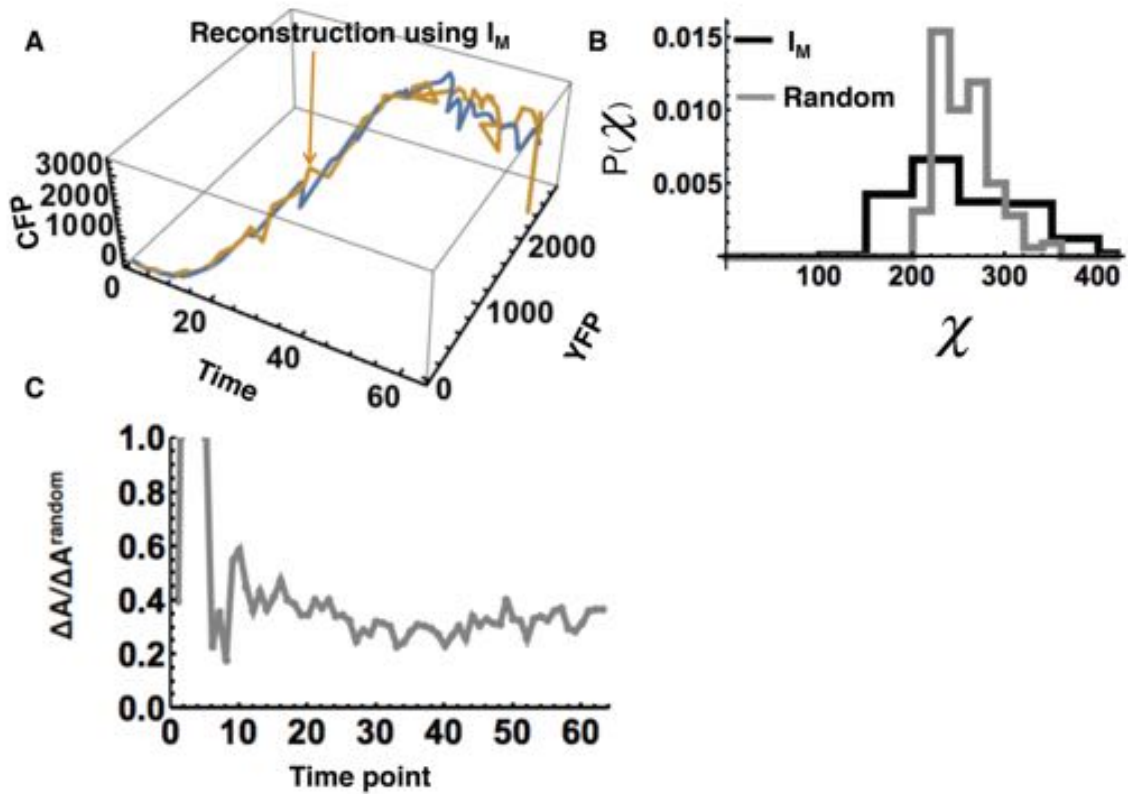

**Figure S12: Quantification of errors in reconstructed trajectories using $I_M$ in live-cell imaging.** The live cell imaging data as described in Fig. 4 was used to generate reconstructed trajectories. We used 159 single cell trajectories for the analysis. **(A)** Shows the reconstructed trajectory using $I_M$. **(B)** Distribution of the error $\chi$, $P(\chi)$ (in black), for the reconstruction carried out using $I_M$. $P(\chi)$ for random pairing is shown in grey for comparison. **(C)** Error in the autocorrelation function in the reconstructed trajectories by $I_M$. The autocorrelation was calculated for the successive pairs of time points (e.g., 0min to 2.5min, 2.5min to 5min, and so on) available in the data. For most of the time points we find $\Delta A/\Delta A_{random} < 1$.

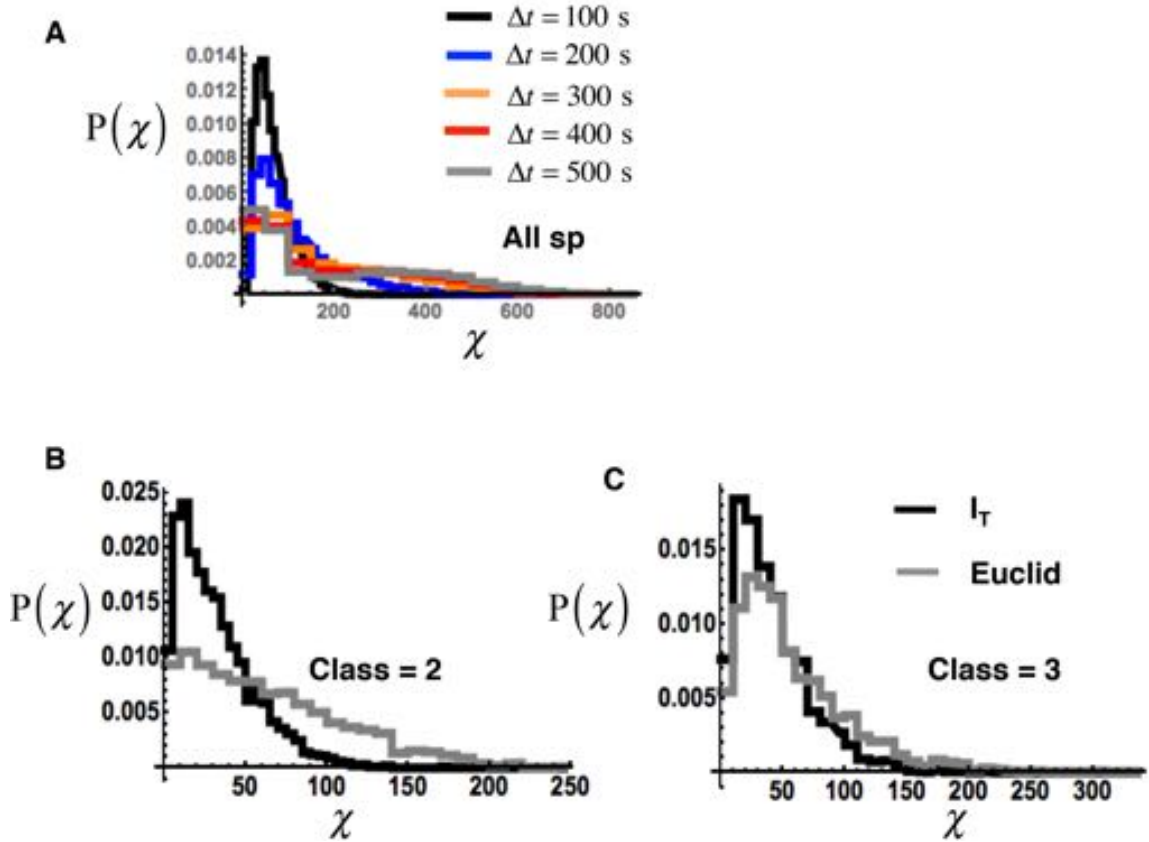

**Figure S13: (A)** Comparison of the distribution of $\chi$ when the time difference between respectively time stamped data for a bistable Ras-SOS network is progressively increased. 6 species namely SOS, RasGDP, RasGTP, RasGAP, DAG and RasGRP were assayed at times t=0 to t=500 s with an increment of 100 s between successive assays. The solid lines in black, blue, orange, red and gray show P($\chi$) for the reconstructions for data sets assayed at time t=0 and t= 100s, t=0 and t=200 s, t=0 and t=300s, t=0 and t=400 s and t=0 and t=500 s respectively. The system approach the bistable Ras activation close to t=300s. **(B)** Shows the distribution of the reconstruction error $\chi$ carried out for the subset that produced minimum $JSD^{(I_T)}$ for k=2 for the data set used in Fig. 6 in the main text. The reconstructions were carried out using $I_T$ (black) or with a method that minimizes the total Euclidean distance (grey). **(C)** Same as (B) except the comparison is done for the subset with minimum $JSD^{(I_T)}$ for k=3.

**Table S1A: Rate constants for the network in Fig. S1**

| Rate constants | min$^{-1}$ |
| --- | --- |
| $k_1$ | 0.08 |
| $k_2$ | 0.1 |
| $k_3$ | 0.145 |
| $k_4$ | 0.226 |
| $k_5$ | 0.321 |
| $k_6$ | 0.178 |
| $k_7$ | 0.134 |
| $k_8$ | 0.245 |
| $k_9$ | 0.48 |
| $k_{10}$ | 0.50 |
| $k_{11}$ | 0.033 |
| $k_{12}$ | 0.23 |
| $k_{13}$ | 0.128 |
| $k_{14}$ | 0.67 |
| $k_{15}$ | 0.45 |
| $k_{16}$ | 0.51 |
| $k_{17}$ | 0.11 |
| $k_{18}$ | 0.05 |
| $k_{19}$ | 0.32 |
| $k_{20}$ | 0.20 |

**Table S1B: Initial conditions for the network in Fig. S1**

The initial species copy numbers are drawn from a multivariate Gaussian distribution. Below we quote the average values and covariance matrix for the species abundances.

| Average | # of molecules |
|---|---|
| Species 1 | 122 |
| Species 2 | 186 |
| Species 3 | 192 |
| Species 4 | 259 |
| Species 5 | 101 |
| Species 6 | 268 |
| Species 7 | 176 |
| Species 8 | 196 |
| Species 9 | 209 |
| Species 10 | 158 |
| Species 11 | 173 |
| Species 12 | 225 |
| Species 13 | 286 |
| Species 14 | 202 |

Covariance Matrix

```
 1326.54  -32.6209  441.557   521.367   23.646   283.172  -143.846  314.964   222.966   218.515  -61.7517  486.626  -594.715  -420.437
 -32.6209 3153.51   65.5223  -84.8321 -12.8702  810.822  -267.412  679.462   439.157  -274.004  30.4547  -62.6026 -219.974   301.26
 441.557   65.5223  3376.05  -29.3363  224.051  938.333  -377.637   52.7194 -250.807   83.6003 -318.352 -367.788    8.1994   13.5782
 521.367  -84.8321  -29.3363 5902.05    44.8606  -7.75098 579.177  -579.456  488.745  349.197  -801.413 -986.762  1479.98    899.332
  23.646  -12.8702  224.051   44.8606  939.701  -496.84   74.7393  167.398  -152.286  233.251  131.992   424.503 -212.305    88.9822
 283.172  810.822  938.333   -7.75098 -496.84   6374.07  -592.042 -194.304  307.236   107.392  -298.845  669.602   657.67   -198.722
-143.846 -267.412 -377.637  579.177    74.7393 -592.042 2738.85  -306.283  -396.695  139.609   381.185  -39.9622 -147.113  -164.373
 314.964  679.462   52.7194 -579.456  167.398  -194.304 -306.283  3436.46   730.01    474.362  539.552  -289.618 -1040.05  -422.465
 222.966  439.157 -250.807  488.745  -152.286  307.236  -396.695  730.01   3926.89    33.2891   71.2203 -338.489 -753.413   421.906
 218.515 -274.004  83.6003  349.197   233.251  107.392  139.609   474.362   33.2891  2246.67  -145.769  440.031  -97.3797   342.974
 -61.7517  30.4547 -318.352 -801.413  131.992 -298.845  381.185   539.552   71.2203 -145.769  2733.75   655.68  -855.245    554.293
 486.626  -62.6026 -367.788 -986.762  424.503  669.602  -39.9622 -289.618 -338.489   440.031   655.68  4528.02    788.084  -636.284
-594.715 -219.974   8.1994  1479.98  -212.305  657.67  -147.113 -1040.05  -753.413  -97.3797 -855.245   788.084  7414.02    951.054
-420.437  301.26   13.5782   899.332   88.9822 -198.722 -164.373 -422.465   421.906  342.974   554.293 -636.284   951.054  3677.34
```

**Table S2A: Rate constants for the Ras activation kinetics network in Fig. S6**

| Rate Constants | |
|---|---|
| $k_1$ | 0.0053 µM$^{-1}$s$^{-1}$ |
| $k_{-1}$ | 4.0 s$^{-1}$ |
| $k_{1cat}$ | 0.0005 s$^{-1}$ |
| $k_2$ | 0.12 µM$^{-1}$s$^{-1}$ |
| $k_{-2}$ | 3.0 s$^{-1}$ |
| $k_3$ | 0.11 µM$^{-1}$s$^{-1}$ |
| $k_{-3}$ | 0.4 s$^{-1}$ |
| $k_4$ | 0.07 µM$^{-1}$s$^{-1}$ |
| $k_{-4}$ | 1.0 s$^{-1}$ |
| $k_{4cat}$ | 0.003 |
| $k_5$ | 0.05 µM$^{-1}$s$^{-1}$ |
| $k_{-5}$ | 0.1 s$^{-1}$ |
| $k_{5cat}$ | 0.038 |
| $k_6$ | 1.74 µM$^{-1}$s$^{-1}$ |
| $k_{-6}$ | 0.2 s$^{-1}$ |
| $k_{6cat}$ | 0.1 s$^{-1}$ |
| $k_7$ | 0.1 µM$^{-1}$s$^{-1}$ |
| $k_{-7}$ | 5.0 s$^{-1}$ |
| $k_8$ | 0.33 µM$^{-1}$s$^{-1}$ |
| $k_{-8}$ | 1.0 s$^{-1}$ |
| $k_{8cat}$ | 0.01 s$^{-1}$ |

Cytosolic volume used =0.08 µm$^3$, surface area and the depth of plasma membrane used are 4.0 µm$^2$ and 1.7 nm respectively.

**Table S2B: Copy numbers for the species used in the Ras activation network in Fig. S6**

The initial copy numbers of the molecular species are drawn from a multivariate Gaussian distribution. Below we show the average values and the covariance used. The species that are not shown here have zero abundances at t=0.

| Average | # of molecules |
|---|---|
| #1. Sos | 85 |
| #2. RasGDP | 370 |
| #3. RasGTP | 30 |
| #4. RasGap | 10 |
| #5. RasGRP1 | 47 |
| #6. DAG | 40 |

Covariance matrix.

$$\begin{pmatrix} 634.324 & 534.672 & 10.0372 & 12.9434 & 50.9679 & 41.6175 \\ 534.672 & 12269.9 & 115.168 & 14.3392 & 343.09 & -111.897 \\ 10.0372 & 115.168 & 79.1503 & 4.11305 & 21.4583 & 9.31962 \\ 12.9434 & 14.3392 & 4.11305 & 9.10811 & -6.54498 & -3.41377 \\ 50.9679 & 343.09 & 21.4583 & -6.54498 & 209.92 & -12.0996 \\ 41.6175 & -111.897 & 9.31962 & -3.41377 & -12.0996 & 138.968 \end{pmatrix}$$

**Pseudocodes for the algorithms used for implementing the framework**

**Notations**
N: Number of species measured
$k \subset N$ {k=2, 3, ..., N}: Class with $k$ species
$m$ {m=1, 2, 3, ..., $^NC_k$}: sub-modules with $k$ species (elements of class $k$)
$[D_{t1}^{m,k}]_{ij}$ {$i$=1, 2, ....., # of cells}, {$j$=1, 2, .., $k$}: Data matrix at time $t_1$ of the $m^{th}$ sub-module belonging to class $k$. For example $[D_{t1}^{1,2}]_{ij}$ contains first 2 species at time $t_1$ whereas $[D_{t1}^{2,2}]_{ij}$ contains the first and the third species at time $t_1$.
$[D_{t2}^{m,k}]_{ij}$ {$i$=1, 2, ....., # of cells}, {$j$=1, 2, .., $k$}: Same as $[D_{t1}^{m,k}]_{ij}$ except the species abundances are measured at time $t_2$.
$[A_{t2}^{m,k}]_{ij}$ {$i$=1, 2, ....., # of cells}, {$j$=1, 2, .., $k$}: Data matrix for the sub-module $m$ belonging to the class $k$ at time $t_2$ that contains the species abundance of the correct partners of the cells in $[D_{t1}^{m,k}]_{ij}$. So if correctly aligned the first cell (row #1) in $[D_{t1}^{m,k}]_{ij}$ would be paired to the first cell (row #1) of $[A_{t2}^{m,k}]_{ij}$.
$M^{JSD}_{ij}$ {$i$=1, 2, ..., N-1}, {$j$=1, 2}: Array for the min JSD values for each class $k \in$ {2, 3, ..., N}.

$[I_T^{t1}]_i = \sum_{j=1}^{k} [D_{t1}^{m,k}]_{i,j}$ {$i$=1, 2, # of cells}: Array containing sum of the species abundances of the $m^{th}$ sub-module belonging to class $k$ at time $t_1$.

$[I_T^{t2}]_i = \sum_{j=1}^{k} [D_{t2}^{m,k}]_{i,j}$ {$i$=1, 2, # of cells}: Array containing sum of the species abundances of the $m^{th}$ sub-module belonging to class $k$ at time $t_2$.

$B_{ij}$ {$i$=1, 2, ...., # of cells}, {$j$=1, 2}: Alignment matrix. If the first row of B is {3, 4} then it means that the $3^{rd}$ cell at time $t_1$ ($3^{rd}$ row in $[D_{t1}^{m,k}]$) is aligned to the $4^{th}$ cell at time $t_2$ ($4^{th}$ row in $[D_{t2}^{m,k}]$).
$QAlg_i$ {$i$=1,N-1}: Array for the quality of alignment

**Algorithm for minimum JSD**

Input: $D_{t1}^{m,k}$, $D_{t2}^{m,k}$

Output: $M^{JSD}$

Initialize $M^{JSD}_{i,1} = \alpha$ where $\alpha > \ln 2 \ \forall \ i \in$ {1, N-1}.
for $k \leftarrow 2$ to $N$ do
    for $m \leftarrow 1$ to $^NC_k$ do
        load $D_{t1}^{m,k}$ and $D_{t2}^{m,k}$
        calculate $I_T^{t1}$ and $I_T^{t2}$
        calculate P= distribution of $I_T^{t1}$ and
            Q= distribution of $I_T^{t2}$
        calculate JSD(P, Q) using eqn (7) main text
        If JSD(P, Q) < $M^{JSD}_{k-1,1}$ then

$$M^{JSD}{}_{k-1,1} = JSD(P, Q) \text{ and } M^{JSD}{}_{k-1,2} = m$$
    endif
   end
 end

Print $M^{JSD}$

**Algorithm for calculation of quality of reconstruction**

Input: $M^{JSD}$, $D_{t1}^{m=m^*,k}, D_{t2}^{m=m^*,k}, A_{t2}^{m=m^*,k}$ where $m^*$ denotes the sub-module with minimum JSD
Ouput: QAlg

for $k \leftarrow 2$ to $N$ do
  Load $D_{t1}^{m^*,k}, D_{t2}^{m^*,k}, A_{t2}^{m^*,k}$ using $M^{JSD}$
 for $i \leftarrow 1$ to # of cells do
   build matrix T1 and T2 such that

$$T1_{i,1} = i \text{ and } T1_{i,2} = \sum_{j=1}^{k}\left[D_{t1}^{m^*,k}\right]_{ij} \text{ and }$$

$$T2_{i,1} = i \text{ and } T2_{i,2} = \sum_{j=1}^{k}\left[D_{t2}^{m^*,k}\right]_{ij}$$

 end

 build ST1 = Sort T1 by column 2
   ST2 = Sort T2 by column 2

 sum1 = 0

 sum2 = 0

 for $i \leftarrow 1$ to # of cells do

   $B_{i,1} = ST1_{i,1}$
   $B_{i,2} = ST2_{i,1}$

$$\text{sum1} = \text{sum1} + \sqrt{\sum_{j=1}^{k}\left(A_{B_{i,1},j} - \left[D_{t2}^{m^*,k}\right]_{B_{i,2},j}\right)^2}$$

$$\text{sum2} = \text{sum2} + \sqrt{\sum_{j=1}^{k}\left(A_{B_{i,1},j} - \left[D_{t2}^{m^*,k}\right]_{Rand[1,\# \text{ of cells}],j}\right)^2}$$

 end

$QAlg_{k-1}$=sum1/sum2
end

**Physical interpretation of the invariant I$_M$**

The time evolution changes the abundance vector $\vec{x}^{(\alpha)}(t)$ as

$$x_i^{(\alpha)}(t_2) = \left[e^{M(t_2-t_1)}\right]_{ij} x_j^{(\alpha)}(t_1) = A_{ij}(t_2,t_1)x_j^{(\alpha)}(t_1) \tag{S1}$$

$A$ is an n×n real matrix. Suppose, the vectors, $\vec{x}^{(\alpha)}(t_2)$ and $\vec{x}^{(\alpha)}(t_1)$, are scaled by matrices S$_2$ and S$_1$, respectively, such that the change given by Eq. (S1) transforms the scaled vectors (denoted by, $\vec{\tilde{x}}^{(\alpha)}(t_2)$ and $\vec{\tilde{x}}^{(\alpha)}(t_1)$, respectively) by an orthogonal matrix, i.e., the scaled vectors undergo rotation or reflection. The scaled vectors are given by,

$$\vec{\tilde{x}}^{(\alpha)}(t_2) = S_2 . \vec{x}^{(\alpha)}(t_2) \tag{S2}$$
$$\vec{\tilde{x}}^{(\alpha)}(t_1) = S_1 . \vec{x}^{(\alpha)}(t_1)$$

The "·" operation in the above equations denotes matrix multiplication. Eq. (S1) can be written in terms of the scaled variables as,

$$\vec{\tilde{x}}^{(\alpha)}(t_2) = S_2 . A . S_1^{-1} . \vec{\tilde{x}}^{(\alpha)}(t_1) \tag{S3}$$

The matrix, S$_2$A can be polar decomposed(1) as,

$S_2 A = QP$
where, Q is an orthogonal matrix (QQ$^T$=I) and
$P = \sqrt{(S_2 A)^T (S_2 A)}$ .

Below we show that when S$_2$=[J(t$_2$)]$^{-1/2}$ =J$_2^{-1/2}$ and S$_1$=[J(t$_1$)]$^{-1/2}$=J$_1^{-1/2}$, then S$_2$AS$_1^{-1}$ in Eq. (S3) is equal to the orthogonal matrix Q.

$$S_2 . A . S_1^{-1} = QPS_1^{-1} = Q\sqrt{(S_2 A)^T (S_2 A)} S_1^{-1} = Q\sqrt{(J_2^{-1/2} A)^T (J_2^{-1/2} A)} J_1^{1/2} = Q\sqrt{(A^T J_2^{-1} A)} J_1^{1/2}$$

According to Eq. (15b) in the main text, J$_2$=AJ$_1$A$^T$ or J$_2^{-1}$=(A$^T$)$^{-1}$ (J$_1$)$^{-1}$A$^{-1}$ or A$^T$ J$_2^{-1}$ A=(J$_1$)$^{-1}$. As a result,

$$S_2 . A . S_1^{-1} = Q\sqrt{(A^T J_2^{-1} A)} J_1^{1/2} = QJ_1^{-1/2} J_1^{1/2} = Q .$$

Therefore, scaling the abundance vector (Eq. (S2) or Eq. (3) in the main text) makes the time evolution of the scaled vectors either a rotation or a reflection. As a result the magnitude of the scaled vector (or I$_M$) does not change with time (Eq. (6) in the main text).

1. Halmos PR (1987) *Finite-dimensional vector spaces* (Springer-Verlag, New York) pp viii, 199 p.